\documentstyle[11pt,psfig,emulateapj]{article}
\def\gsimeq
{\hbox{\raise0.5ex\hbox{$>\lower1.06ex\hbox{$\kern-1.07em{\sim}$}$}}}
\def\lsimeq
{\hbox{\raise0.5ex\hbox{$<\lower1.06ex\hbox{$\kern-1.07em{\sim}$}$}}}
\def\pn{\par\noindent}
\def\ss{\smallskip\pn}
\def\ms{\medskip\pn}

\begin{document}

\renewcommand{\thefootnote}{\fnsymbol{footnote}}

\title{$Chandra$ study of an overdensity of X-ray sources around two
distant (z$\sim$0.5) clusters} 

\author{M. Cappi$^{1,2}$, P. Mazzotta$^1$, M. Elvis$^1$, D.J. Burke$^3$, A. Comastri$^4$, \\
F. Fiore$^5$, W. Forman$^1$, A. Fruscione$^1$, P. Green$^1$, D. Harris$^1$, \\
E.J. Hooper$^1$, C. Jones$^1$, J.S. Kaastra$^6$, E. Kellogg$^1$, S. Murray$^1$, B. McNamara$^1$, \\
F. Nicastro$^{1,5}$, T.J. Ponman$^{1,7}$, E.M. Schlegel$^1$, A. Siemiginowska$^1$, \\
H. Tananbaum$^1$, A. Vikhlinin$^{1}$, S. Virani$^1$, B. Wilkes$^1$ \\}


\small
\begin{center}
\pn
   $^1$ Harvard-Smithsonian Center for Astrophysics, 60 Garden St., Cambridge, MA 02138, USA
\pn
   $^2$ Istituto TeSRE-CNR, Via Gobetti 101, 40131, Bologna, Italy
\pn
   $^3$ Institute for Astronomy, University of Hawaii, 2680 Woodlawn Drive, Honolulu, HI 96822
\pn
   $^4$ Osservatorio Astronomico di Bologna, Via Ranzani 1, 40127, Bologna, Italy
\pn
   $^5$ Osservatorio Astronomico di Roma, Via dell'Osservatorio, I-00044 MontePorzio Catone, Italy
\pn
   $^6$ SRON, Sorbonnelaan 2, NL-3584 CA Utrecht, The Netherlands
\pn 
   $^7$ School of Physics \& Astronomy, University of Birmingham, Birmingham B15 2TT, UK
\end{center}
\normalsize


\begin{abstract}
We present results from a $Chandra$ X-ray Observatory study
of the field X-ray source populations in 4 different observations: 
two high-redshift (z$\sim$0.5) clusters of 
galaxies 3C295 and RXJ003033.2+261819; and two non-cluster fields with 
similar exposure time. 
Surprisingly, the 0.5-2 keV source surface densities
($\sim$ 900-1200 sources deg$^{-2}$ at a flux limit 
of 1.5$\times$ 10$^{-15}$ erg cm$^{-2}$s$^{-1}$) measured in an 
$\sim$8$^\prime$$\times$8$^\prime$ area 
surrounding each cluster exceed by a factor of $\sim$ 2 the value 
expected on the basis of the {\it ROSAT} and {\it Chandra} logN-logS, with a significance of 
$\sim 2 \sigma$ each, or $\sim$ 3.5 $\sigma$ when the 2 fields are combined 
(i.e. a probability to be a statistical fluctuation of 
$<$1\% and $<$0.04\%, respectively). The same analysis 
performed on the non-cluster fields and on the outer chips of the cluster 
fields does $not$ show evidence of such an excess.
In both cluster fields, the summed 0.5-10 keV spectrum of the detected objects 
is well fitted by a power-law with $\Gamma$ $\sim$ 1.7 similar to AGNs 
and shows no sign of intrinsic absorption. 
The few ($\sim$ 10 out of 35) optical identifications available to date confirm that most 
of them are, as expected, 
AGNs but the number of redshifts available is too small to allow conclusions 
on their nature.
We discuss possible interpretations of the overdensity in terms of: 
a statistical variation of Cosmic Background sources; a  
concentration of AGNs and/or powerful starburst galaxies associated with the 
clusters; and gravitational lensing of background QSO's by the 
galaxy clusters. All explanations are however difficult to reconcile 
with the large number of excess sources detected.
Deeper X-ray observations and more redshifts measurements are 
clearly required to settle the issue.

\noindent {\em Subject Headings:} galaxies: active:clustering:general - X-rays:general

\end{abstract}

\section{Introduction}

Since its launch date on July 23rd, 1999 the $Chandra$ X-ray Observatory
has performed a number of pointed observations aimed at verifying the 
satellite functioning and at calibrating the instrument responses.
This paper reports the analysis of serendipitous sources detected 
in three of these observations.

Among the most remarkable characteristics of $Chandra$ are 
its unprecedented sensitivity and spatial resolution ($\lsimeq$1 arsec) 
over the entire 0.1-10 keV band (Van Speybroeck, et al. 1997), 
a factor of $\sim$ 10 better than any previous X-ray mission. 
This provides an order-of-magnitude advance in
detecting faint point sources ($\sim$ 10--100 times fainter than {\it ROSAT} and
{\it ASCA} at a given exposure time) because of the 100 times reduced background 
per beam element. 
On the basis of the {\it ROSAT} measurements (Hasinger et al. 1998),
at a 0.5-2 keV flux limit of 
3$\times$10$^{-15}$ erg~cm$^{-2}$~s$^{-1}$, the source density in $Chandra$
observations is expected to be 340$\pm$30 deg$^{-2}$, giving $\sim$ 6 
sources per chip. This opens new possibilities for detecting many 
serendipitous X-ray sources even in
observations with modest ($\sim$ few tens ks) exposures. 
This will lead to the collection of sufficiently large samples to enable
detailed study of the logN - logS, the X-ray background, and potentially
the spatial distribution of sources to map large-scale structure.

In this paper, we report on the serendipitous sources in 
{\it Chandra} observations of two 
medium-z clusters RXJ003033.2+261819 (z=0.5, Vikhlinin et al. 1998, 
``RXJ0030'' hereinafter) and 3C295 (z=0.46, Dressler \& Gunn 1992) which suggest an excess
number of serendipitous X-ray sources compared to a non-cluster field and 
to the predictions based on the {\it ROSAT} and {\it Chandra} (0.5-2 keV) logN-logS measurements.
The clusters have 0.5-10 keV luminosities of $\sim$ 10$^{44}$ and 10$^{45}$ erg s$^{-1}$ 
and temperatures of $\sim$ 4 and 4.4 keV, respectively (Forman et al. 2000, in preparation, 
Harris et al. 2000).
We first show (\S 4.1) the source densities obtained from the on-axis chips 
of the RXJ0030 and 3C295 fields of view (FOVs), and compare them to 
densities in the outer chips and to two comparison fields
(\S 4.2 and 4.3): one obtained when $Chandra$ pointed away from 
the radiant of the 1999 Leonid meteor shower (hereafter anti-Leonid) and 
one obtained from a calibration observation of 3C273.
The average X-ray spectral properties and available optical identifications 
of these sources are given in \S 5.
Possible interpretations of these results are discussed 
in \S 6 and conclusions are reported in \S 7.
$H_0=50$ km s$^{-1}$ ~Mpc$^{-1}$ and $q_0=0.5$ are used throughout.

\section{Observations and Data Reductions}

The {\it Chandra} X-ray Observatory (Weisskopf, O'dell \& van Speybroeck, 1996) 
consists of four pairs of concentric Wolter I mirrors reflecting 
0.1-10 keV X-rays (VanSpeybroeck, et al. 1997) into one
of the four focal plane detectors (ACIS-I/S or HRC-I/S).
All the data presented in the following were taken from the 
$Chandra$ public archive (Fabbiano et al. 2000, in preparation, 
see also http://asc.harvard.edu/cda/).
The observations of RXJ0030 and 3C295 were performed with the
ACIS-S, with the clusters lying within a few arcsecs of the optical axis 
location on the back-illuminated (BI) S3 chip. Chips S1-S4, I2 and I3
constitute the entire activated field of view.
The anti-Leonid observation was 
performed with the ACIS-I configuration (i.e. with the focus nearly 
at the center of four front-illuminated (FI) CCDs, 
see the ``{\it Chandra} Proposer's Observatory Guide'' 1999).
The observation of 3C273 was performed with a ACIS-S(1-6) configuration.
These two observations were chosen from the public archival data 
as the best available comparison fields because 
of their long exposures and high Galactic latitudes.

Details of the cleaning and reduction of the data are given in Appendix I.
In total, we obtained ``good'' 0.1-10 keV data from 4 chips for each FOV, 
each chip having dimensions of 8$^{\prime} \times 8^{\prime}$.
A log of the observations is given in Table 1.

\section{Analysis: The Source Detections, Counts and Fluxes}

\subsection{The Source Detections}

To localize the serendipitous source candidates in the fields,
we applied a source detection algorithm in the $Chandra$ Interactive Analysis of 
Observations (CIAO, Elvis et al. 2000, in preparation) software: 
$wavdetect$ (Freeman et al. 2000, Dobrzycki et al. 1999). 
Source detection is easy with $Chandra$ because 
the background is very uniform and low, even in the vicinity of the clusters,
and one can therefore ``see'' the sources unambiguously. With the conservative 
threshold applied here (see Appendix II), all sources are indeed visible 
by eye. Details of the algorithm and the procedure applied for the detections 
are given in Appendix II.

In the 0.5-2 keV energy band, a total of 53 and 44
point sources/4-chip FOV (clusters excluded) are found 
in the RXJ0030 and 3C295 fields, respectively. 
All detected sources were consistent with point sources.
More sources are expected in the RXJ0030 field because we have 
about 50\% more usable exposure time.
In the hard 2-10 keV energy band, these numbers are reduced to 
13 and 5, respectively. If we restrict
ourselves to the central S3 chip (where systematic effects are expected
to be smaller), we find 23 and 17 sources 
between 0.5-2 keV, and 6 and 4 sources between 2-10 keV,
respectively. These sources are listed in Tables A.1 and A.2.
For comparison, we also list the sources detected in 
the whole 4 chips of the anti-Leonid field and in the ACIS-S3 chip 
of the 3C273 field (Tables A.3 and A.4).

Figure 1 shows an overlay of the detected sources with the
X-ray images between 0.5-2 keV, where the images have
been smoothed using a gaussian function with $\sigma$ = 1 pixel = 2''.
All the detected sources are clearly visible in the images.

\subsection{The Source Counts and Fluxes}

Source count rates were obtained using the $wavdetect$ 
algorithm from regions with typical radii of $\sim$ 3'' 
(on-axis) and 10'' (off-axis).
The measured counts were first corrected for vignetting
and then converted to an emitted, unabsorbed flux. 
A description of this procedure is given in Appendix III.

Tables A.1 and A.2 report the 0.5-2 keV and 2-10 keV measured fluxes for the 
detected sources (column 6) and, for sources with known redshifts, 
the corresponding luminosities.
The total 0.5-2 keV flux of the point sources in RXJ0030 (most of which are within 
a 5 arcmin radius) is 2.2 $\times$ 10$^{-13}$ ergs cm$^{-2}$ s$^{-1}$, 
$\sim$ 1.6 times larger than the cluster flux over the same energy band. 
About half of this flux is in the z=0.492 CRSS QSO (source 1 in Table A.1). 
In 3C295, the point sources sum to 1.6 $\times$ 10$^{-13}$ erg cm$^{-2}$ s$^{-1}$, 
about one third of the cluster flux.

\section{Densities of the Serendipitous Sources}

\subsection{In the Central Chips (S3)}

Here we focus on the results obtained from the central chip (S3) of the RXJ0030
and 3C295 fields, which contain the clusters themselves, and compare them to
the {\it ROSAT}, {\it ASCA} and {\it Chandra} logN-logS. 

The faintest source in the RXJ0030 field has a flux of 
$\sim$ 1.3$\times$10$^{-15}$ erg~cm$^{-2}$~s$^{-1}$ (0.5-2 keV) and $\sim$ 10$^{-14}$
erg~cm$^{-2}$~s$^{-1}$ (2-10 keV). For 3C295, the faintest has $\sim$ 1.3 $\times$ 10$^{-15}$
erg~cm$^{-2}$~s$^{-1}$ and $\sim$ 3.3 $\times$10$^{-14}$ erg~cm$^{-2}$~s$^{-1}$ 
between 0.5-2 and 2-10 keV, respectively. All the sources detected at 
2-10 keV were also detected in the 0.5-2 keV band.

We find that the minimum number of photon counts of the detected 
sources increases only very weakly as the off-axis distance increases. 
This is especially true if one considers only the central chip S3 (where the
off-axis distance is confined to $\lsimeq$ 7 arcmin). 

At a flux limit of 1.5 $\times$10$^{-15}$ erg~cm$^{-2}$~s$^{-1}$, 
a source gives $\sim$ 15 and 10 net counts per detected source in RXJ0030 and 3C295, 
compared with a background of $\sim$ 3 and 1 counts per source extraction 
area, respectively. Hence most of the sources should have been detected 
regardless of position in the FOV.
Given the negligible background most of the sources 
have a signal-to-noise ratio higher than 4 at this flux limit, and we therefore 
avoid complications due to ``Eddington bias'', as shown by 
Schmitt \& Maccacaro (1986).


The main results of the present study are given in Table 2 and are shown in 
Figures 2 and 3.
The table reports the source densities per deg$^{2}$ measured in the central chip 
(S3) of RXJ0030 and 3C295 for two different
flux limits (1.5 and 3 $\times$ 10$^{-15}$ erg~cm$^{-2}$~s$^{-1}$) for the 0.5-2 keV
detections and for 2 $\times$ 10$^{-14}$ erg~cm$^{-2}$~s$^{-1}$ for the 2-10 keV
detections. 
For comparison, we also report in the same table the logN-logS obtained from 
the {\it ROSAT} Lockman Hole deep-field (Hasinger et al. 1998), two recent 
$Chandra$ deep-fields (Mushotzky et al. 2000; Giacconi et al. 2000) and from 
the two comparison fields.
The logN-logS curves (upper panel of Figure 2) show the Table 2 numbers 
plus those obtained at fluxes 
of 2.5 and 4 $\times$ 10$^{-15}$ erg~cm$^{-2}$~s$^{-1}$. 
We used the geometric area of 1 chip (64 arcmin$^2$) for RXJ0030, 3C295 and 3C273, and 
4 chips (256 arcmin$^2$) for the anti-Leonid field. These are upper-limits on the real area given 
the uncertainties in instrumental effects (see Appendix II).

It is clear from Table 2a and Figure 2 that the present 
0.5-2 keV results strongly suggest an excess number of serendipitous sources in 
the RXJ0030 and 3C295 fields when compared to {\it ROSAT} and {\it Chandra} deep-field 
counts. 
Assuming the {\it ROSAT} value as the true mean source density, the probability of 
finding a number of sources equal or greater than that observed in RXJ0030 is 
1\% and 0.4\% for fluxes of 3 and 1.5$\times$10$^{-15}$ erg~cm$^{-2}$~s$^{-1}$, 
respectively. For 3C295, the probabilities are 1\% and 8\% for the same fluxes.

If the allowed 1$\sigma$ upper envelope on the {\it ROSAT} measurements are used instead, 
the probabilities are about two times higher. At a higher flux limit of 
2.5$\times$10$^{-15}$ erg~cm$^{-2}$~s$^{-1}$, the probability is even 
lower for RXJ0030.
Based on measurements of the angular correlation 
function of X-ray sources derived from a {\it ROSAT} PSPC survey by 
Vikhlinin \& Forman (1995), we estimate that the \emph{rms} fluctuations 
of the X-ray source density are on average $\sim$20-35\%. Hence, 
even if cosmic variance is taken into account (see \S6.1.1), 
the probabilities remain similar.
In summary, the excess is significant at the $\sim$ 2 $\sigma$ level in both cases,
even when the uncertainties in the {\it ROSAT} estimates are considered.
If the two fields are combined, then the probabilities at 
1.5 and 3$\times$10$^{-15}$ erg~cm$^{-2}$~s$^{-1}$ become 0.1\% and 0.04\%, i.e. a 
significance of $\sim$ 3 and 3.5 $\sigma$ which is larger at higher than lower 
fluxes.

In the hard energy band, although we reach a factor 2-3 deeper 
(2$\times$10$^{-14}$ erg~cm$^{-2}$~s$^{-1}$) 
than the {\it ASCA} and {\it BeppoSAX} logN-logS (Giommi, Perri \& Fiore 2000), the 
statistics are poorer and the source counts from the four $Chandra$ fields 
are consistent with the Comastri et al. (1999) model for the XRB, 
the {\it ASCA} fluctuations, and the $Chandra$ deep-field observations (see Figure 3).

Recently, Brandt et al. (2000) identified some of the $Chandra$ 2-8 keV sources in the 
RXJ0030 field (see also \S5.2). 
They detect 9 sources in the central chip. Five out of 6 of our 
sources are coincident with their detections and the remaining 4 are 
below our conservatively chosen threshold.
Brandt et al. do not detect our source 2 in Table A.1 because they exclude 
the chip border. At our hard band flux limit of 2$\times$10$^{-14}$ erg cm$^{-2}$ s$^{-1}$, 
their derived source density is consistent with the one presented here.

\subsection{In the Outer Chips, and in the anti-Leonid and 3C273 Comparison Fields} 

Following the procedure explained in \S 3, we have calculated the 
0.5-2 keV logN-logS distribution of the sources detected 
in the 3 ``external'' chips of RXJ0030 and 3C295 (2 ACIS-I + 1 ACIS-S chip), 
in the full anti-Leonid FOV (4 ACIS-I chips) and in the S3 chip of the 3C273 field.
These are shown in the lower panel of Figure 2.
The results for the anti-Leonid field, the 3C273 field and the source density 
in the outer regions of the cluster fields are fully consistent with 
the {\it ROSAT} and {\it Chandra} logN-logS (note that the Mushotzky 
et al. results were obtained from an observation performed with the 
S3 chip, like the present cluster fields).  
These agreements are important because they demonstrate that the 
excess of sources measured in the central S3 chips of RXJ0030 and 3C295 
(\S4.1) are not due to an instrumental effect.

The possibility that the ``surplus'' sources are caused 
by some statistical fluctuations due to the enhanced background near 
the clusters haloes can be rejected. We find no trend 
for a higher background (computed locally) around the sources nearer 
to the clusters (see Tables A.1 and A.2) and a similar effect would 
be expected in the 3C273 field (compare Fig. 1a,b with Fig. 1d), 
which is not observed.

\subsection{Spatial Distribution}

We computed the density of the sources as a function 
of off-axis distance for all fields.
For the two cluster fields, this clearly corresponds to the 
(angular) distance from the cluster centers since they are 
almost coincident with the optical axis.  
In this case, we made a conservative choice for the flux 
limit of $\sim$ 2.5$\times$10$^{-15}$ erg cm$^{-2}$ s$^{-1}$ 
between 0.5-2 keV. We used this value instead of 1.5 or 
3$\times$10$^{-15}$ erg cm$^{-2}$ s$^{-1}$ given in Table 2 because we 
made a trade-off between limiting as much as possible systematic 
effects at off-axis distance (see \S3) and having the largest 
number of sources. Starting from the center, bins were 
constructed by adding 100'' to its radius until 
there were 9 sources in each bin.
The number of sources/bin were then divided by the area covered by the annuli 
and the resulting source densities plotted as a function 
of off-axis distance (Figure 4). At this flux limit, only 5 sources were 
detected in the 3C273 S3 chip, resulting in a single bin being plotted in Figure 4.
This figure suggests, again, that the source densities are 
higher by a factor $\sim$ 2 near the cluster centers than in the outer regions, 
while there is no radial dependence in the anti-Leonid comparison field. 
Moreover, at off-axis distances larger than $\sim$ 200 arcsec 
(corresponding to $\sim$ 1.4 Mpc at z=0.5 for 
$H_0=50$ km s$^{-1}$ ~Mpc$^{-1}$ and $q_0=0.5$), the source density 
in all fields is consistent with the {\it ROSAT} (Hasinger et al. 1998) 
and {\it Chandra} (Mushotzky et al. 2000; Giacconi et al. 2000) 
logN-logS.

\section{Properties of the Serendipitous Sources}

\subsection{Summed X-ray Spectra}

Here we present the results obtained from the
spectral analysis of the co-added spectrum from all the S3 sources in the
RXJ0030 and 3C295 central fields (i.e. the average spectra of sources 
listed in Tables A.1 and A.2).
We also present the summed spectrum of all serendipitous sources in the 
two comparison fields.
For both fields, data obtained from the brightest of the sources 
(the first source in Tables A.1-4) have been excluded from the summed spectra.


A single spectrum has been constructed that includes the sum of all
counts extracted from elliptical regions (chosen to match the spatially varying 
$Chandra$ PSF) centered on the detected sources, with minor and major axis
typically between about 3 and 7 arsec, chosen to ensure inclusion of more than
90\% of the PSF encircled energy, at all energies and off-axis 
distances. Pulse invariant (PI) response matrices released in October 1999 were 
used. 
We used only the data between 0.5-10 keV, where the matrices are best 
calibrated. The charge transfer inefficiency problems of the 
FI CCD (for the anti-Leonid field only) are not corrected for; changes can be expected 
once more calibrations become available.
The spectra were fitted using the {\it Sherpa} fitting and modeling
application included in CIAO. We applied $\chi ^2$ statistics with the 
Gehrels (1986) approximation of errors in the low counts regime. 

A somewhat critical point of this analysis is that spectral 
deviations due to energy-dependent vignetting 
could artificially steepen our averaged spectrum since, as discussed in 
Appendix II, the vignetting is larger at higher energies.
In the cluster fields (of which only the central chips were considered 
for the spectral analysis) and in the 3C273 field, the vignetting should 
not significantly affect our results 
since most of the effective area (and therefore counts) is at 
E$\lsimeq$4 keV where 
the vignetting is negligible within $\sim$ 5$^\prime$. 
Indeed the spectra constructed from only the off-axis sources 
(at $>$ 5$^\prime$ off-axis) are consistent with that of 
the on-axis sources.
In the case of the anti-Leonid field, we have limited the spectral analysis 
to the 24 sources detected within an off-axis distance 
$\lsimeq$ 7$^\prime$ for comparison with the other fields.

Background contributes a significant fraction ($>$50\%) of the counts at E$>$ 4 keV. 
We extracted background from several (more than 3) large
circles of radii $\lsimeq$ 70 arcsec, chosen randomly in regions with
no detected point sources and co-added their spectra.
We also considered a background chosen from annular
regions around each source regions, and again added all the counts into a single 
spectrum. We found that the two background choices gave best-fit parameters
consistent with each other to within $\sim$ 10\%. 
Hereafter, we report the results obtained with
the background determined from the large circles since this has better statistics.

Best-fit results are given in Table 3 and are shown in Figure 5.
The main result of this analysis is that the summed spectra are 
consistent with a single power-law model ($\Gamma \sim 1.7$) with no absorption 
in excess of the Galactic value. The spectra in the RXJ0030 
and 3C295 fields ($\Gamma \simeq 1.7\pm0.2$) are slightly flatter than the 
summed spectrum in the anti-Leonid field ($\Gamma \simeq 2.3 \pm 0.2$) but 
softer than in the 3C273 field ($\Gamma \simeq 1.2 \pm 0.3$). 
the steeper spectrum from the anti-Leonid field could be due 
to a CTI effect that mostly affects the FI chips.

The spectra can also be described by a high-temperature (kT$\sim$2-4 keV 
if z=0 and kT$\sim$4-6 keV if z=0.5) thermal model 
with poorly constrained abundances ($\sim$0.1-1 solar) and with $\chi^2$ values 
comparable to the single power-law model. The two models are 
indistinguishable, which is not surprising given the limited 
statistics.
In addition we note that the summed spectra of sources detected only in the 
soft X-ray band are in all cases consistent with the ``total'' summed spectra, 
in agreement with the fact that we detect no ``hard X-ray only'' sources (\S4.1).

\subsection{Optical Identifications}

We searched for optical counterparts to the 
S3 X-ray sources in RXJ0030 and 3C295
 from the USNO-A2.0 catalog (Monet et al. 1998) which has $1 \sigma$ 
positional errors $\sim 0.25$ arcsec (Deutsch 1999)
and reaches a $B$ magnitude limit of about 20. 
We used a search radius of 3 arcsec. For larger radii, the fraction of 
random matches exceeds 10\%. 
The USNO-A2.0 catalog includes $B_J$ and $R$ magnitudes.
For ease of comparison with other works, we convert optical magnitudes
originally in the $B_J$ band to the $B$ band using $B=B_J+0.28(B-V)$
(Blair \& Gilmore 1982), assuming $(B-V)=(B-R)/2$.  We apply a
correction for extinction using the results of Burstein \& Heiles
(1978, 1982) and assuming $A_B=4.0\,E(B-V)$. 
In addition, $B$ magnitudes, redshifts and classifications were obtained 
from the NED database for 2 objects in RXJ0030 and 5 in 3C295.
We also included the recent identifications by Brandt et al. (2000) of 
3 additional sources in the field of RXJ0030. 
Dressler \& Gunn (1992) mapped the 
central ($\sim$4$^{\prime}\times4^{\prime}$) regions of 3C295 and obtained photometric 
redshifts for more than 100 galaxies in that area including 2 of the $Chandra$ source 
counterparts (Table A). 
In total, we obtain photometric data for 13 sources and redshifts for 
10 of these in the RXJ0030 and 3C295 central chips (Tables A.1 and A.2).

Starting from the $B$ magnitudes and the derived X-ray (0.5-2 keV) fluxes, we 
calculate the nominal X-ray to optical slope $\alpha_{\rm ox}$ following 
Stocke et al. (1991).
$B$ magnitudes, $\alpha_{\rm ox}$, redshifts, absolute $M_{\rm B}$ and 
classification (if available) of the detected sources are reported in Tables A.1 and A.2.


In RXJ0030, four of the (5) sources 
with redshifts are forgeround galaxies while the 
bright CRSS QSO has z=0.492, which makes it a possible cluster 
member. However, only 5 out of 23 sources have redshifts.
Of the 9 sources within 3$^{\prime}$ of 3C295, 5 have redshifts: 
2 are high-z QSOs, 2 are galaxies at z$\sim$0.6, and 
one (row 1 in Table A.2) is a Seyfert 1 associated with the cluster 
($\Delta z =0.01$). The redshifts of the two galaxies have been estimated from 
optical colors and have relatively large errors ($\sim$0.1; 
Thimm et al. 1994) so these two galaxies could be located at 
the cluster redshift. 

Summarising, one source (out of 4 with redshifts) 
in RXJ0030 and three sources (out of 5 with redshifts) in 3C295 are 
possibly associated with the clusters. 
Given the current poor statistics and redshift uncertainties, 
deeper optical imaging and spectroscopy of more of these objects 
are needed to classify the excess sources and to
determine whether they are associated with the clusters.

\section{Discussion}

The $Chandra$ observations discussed in the present paper 
strongly suggest a factor $\sim$ 2 overdensity of 0.5-2 keV X-ray sources 
around two high-z clusters compared to field X-ray sources 
at a significance level of $\sim$2$\sigma$ each, 
or 3.5$\sigma$ when combined.
Given the present statistics, we cannot rule out that we have 
measured (twice) a statistical fluctuation of the population 
of field X-ray sources.
Clearly, further deeper X-ray observations are required 
to confirm the reality of the excess.
It is however notable that this effect has been found around 2 different 
clusters (the only two in the 
$Chandra$ archive that are currently public) and at roughly the same 
significance, making the chance coincidence rather unlikely. 
Our analysis of the outskirts of the FOVs suggests that the 
excess of sources disappears at large distance ($\gsimeq$1.5 Mpc) from the 
clusters.
At least half of these sources must be the sources which produce 
the XRB (possibly a mixture of QSOs and Seyfert-type 
AGNs, Hasinger et al. 1998, Schmidt et al. 1998, Comastri et al. 1999). 
If the remaining ``surplus'' X-ray sources are 
associated with the clusters, then their average 
luminosity is $\sim$ 10$^{42-43}$ erg s$^{-1}$ in either the 0.5-2 keV or 
2-10 keV bands.
The effect merits some discussion.

\subsection{On the Nature of the surplus X-ray sources}

We shall here consider some possible 
explanations for the origin of these ``surplus'' X-ray sources.

\subsubsection{Cosmic Variance?}

Our view of the Universe now includes a web-like network 
of large-scale structures which include galaxies, clusters and 
filaments (Peebles 1993, Peacock 1999). Two-point correlation 
functions have been able to probe mass clustering on scales 
$\lsimeq$ 5 Mpc for galaxies (e.g. Small et al. 1999), $\lsimeq$ 10 Mpc 
for QSOs (Shaver et al. 1984, Kundic 1997, La Franca, 
Andreani \& Cristiani 1998) and $\lsimeq$ 100 Mpc for clusters 
(see Bahcall 1988 for a review).
If X-ray sources are distributed like galaxies, then their 
surface density will have fluctuations due to the large scale structure. 
These fluctuations are known as ``cosmic variance''. 
Could cosmic variance produce the observed effect?
Few studies have been made in X-rays (see Barcons et al. 2000 for a recent review). 
Studies of complete samples of X-ray selected AGNs 
that probe the $\sim$ 1 Mpc scale of interest
have only become possible in the last few years (Boyle \& Mo 1993) 
and have found positive clustering signals on intermediate scales 
(0.5$^{\prime}$--10$^{\prime}$, 
Vikhlinin \& Forman 1995; correlation length $r_c$ $\lsimeq$ 40-80 $h^{-1}$ Mpc, 
Carrera et al. 1998) and smaller scales ($r_c$ $\simeq$ 6 $\pm$ 1.6 $h^{-1}$ Mpc,
Akylas, Georgantopoulos \& Plionis 2000). 
The question here is: Is the present overdensity consistent with 
a random sample of cosmic variance?
The answer is no, since the observed amplitude is 20-30\%, 
i.e. much lower than the factor $\sim$ 2 fluctuations we measure here.
Because we are looking at regions which are centered on distant clusters, 
we have however a highly biased sample.
If a filament near a cluster lies mostly normal 
to the plane of the sky, then a source excess due to AGNs could however be 
produced near to the clusters.
If this is the case, this could well represent the first $direct$ measurement of 
large-scale structure of X-ray selected sources.

\subsubsection{AGNs/quasars associated with the clusters?}

Following Forman et al. (2000, in preparation), we estimate that the virial radii of 
RXJ0030 and 3C295, corresponding to a mean gas overdensity of 180, 
is $\sim$ 1 Mpc (i.e. 3$^\prime$ at z=0.5). It is possible, therefore, that 
the sources are physically associated with the clusters.

All the objects with optical counterparts have $\alpha_{\rm ox}$ consistent with type 1 AGNs 
(Elvis et al. 1994). If placed at the clusters' redshifts 
all sources, including those with optical counterparts, would have X-ray luminosities of 
$L_X(0.5-2{\rm keV}) \gsimeq 10^{42}$ erg s$^{-1}$ and 
$M_B\lsimeq -23$, again consistent with being bright Seyfert~1 type 
AGN or low luminosity QSOs. 

Supporting the AGN hypothesis are: (a) the shape of the summed spectra which are similar to
that commonly seen in Seyfert type 1 galaxies, (b) the large luminosities 
(few $\times$ 10$^{42}$ erg s$^{-1}$, Table 3) and (c) the $\alpha_{\rm ox}$ of the 
few X-ray sources with optical counterparts (Tables A.1. and A.2). All three properties are typical of 
Seyfert galaxies and QSOs making this explanation attractive. 

This would be however quite strange because AGN are normally rare in clusters.
Although the galaxy environment of AGNs 
is not well studied (Krolik 1999) in the local Universe, AGNs are 
known to occur more frequently ($\sim$5\%) in field galaxies than in nearby (z$<$0.1) 
cluster of galaxies ($\sim$1\%, Osterbrock 1960, Dressler, Thompson \& Shectman 1985).
There is no strong increase in the number of optically selected AGNs in typical 
higher-z clusters (e.g. Ellingson et al. 1997, 1998; Dressler et al. 1999).

3C295 and RXJ0030 may not be typical high-z clusters. 
Dressler \& Gunn (1983) and Dressler et al. (1999) measured a frequency 
of AGNs of $\sim$10\% in 3C295, i.e. $\sim$ 10 times larger than in other 
distant clusters (see Table 7 in Dressler et al. 1999). The pioneering X-ray 
work of Henry et al. (1985) also pointed out that the (X-ray selected) 
AGN population of 3C295 is larger than in low-redshift clusters.
In conclusion, the AGN/quasars hypothesis requires 3C295 and RXJ0030 
to be intrinsically unusual.


\subsubsection{Powerful starburst galaxies associated with the clusters?} 

The limited statistics of the spectra of 
these faint X-ray sources do not allow us to place stringent 
constraints on the origin of their X-ray emission. The spectra are equally well 
described by a single power-law model or by a thermal model 
with kT$\sim$few keV (\S 5.1), consistent with 
the spectra of nearby starburst galaxies (Ptak et al. 1997, 
Dahlem, Weaver \& Heckman 1998, Cappi et al. 1999). 
The spatial extents of the X-ray sources compared with the $Chandra$ PSF, 
are consistent with being point like. 
The most stringent cases (on-axis) place a limit of $\lsimeq$ 2'' on the radius of the X-ray 
emitting region which, at z=0.5, corresponds to a poorly constraining 
limit of $\sim$ 15 kpc (starbursts have typical dimensions of 
$\lsimeq$ 2 kpc, e.g. Heckman, Armus \& Miley 1990).
However, this hypothesis appears to be rather unlikely because 
the X-ray luminosities associated with the detected X-ray sources 
would be a factor of $\sim$ 10-100 times larger than usually observed in nearby 
starburst and normal galaxies (Fabbiano, Kim \& Trinchieri 1992, 
David, Jones \& Forman 1992)
Moreover, there is no suggestion in the 
literature for a significant number of starbursts in distant clusters, 
not any more prevalent than in the field (e.g., Balogh et al. 1999). 

Recent {\it Chandra} deep-surveys 
(Hornschemeier et al. 2000; Mushotzky et al. 2000, Fiore et al. 2000) 
have nevertheless found X-ray bright galaxies at similar redshifts 
with luminosities as high as 
a {\it few} $\times$ 10$^{42}$ erg s$^{-1}$ that could either 
be ellipticals or star-forming galaxies.
And as discussed in the previous section, 3C295 and RXJ0030 could 
be unusual. Indeed, the 3C295 cluster was first picked out 
(Butcher \& Oemler 1978) as one of the first clusters 
discovered to have a larger percentage of blue galaxies than in nearby 
clusters (``Butcher-Oemler effect'', but see Dressler \& Gunn 1983). 
The colors of the blue galaxies suggest they are probably 
undergoing star-formation; they may indeed be 
powerful starbursts in spiral galaxies surrounding the cluster 
(Dressler \& Gunn 1983; Poggianti et al. 1999).

\subsubsection{Gravitationally lensed sources?}

Most rich clusters produce weakly distorted images of background 
galaxies/QSOs in the optical band (weak lensing) and occasionally giant arcs 
(strong lensing).
Both 3C295 (Smail et al. 1997) and RXJ0030 (B. McNamara, 
private communication) show evidence of such effects implying cluster masses of 
the order of $\sim$ 10$^{14}$M$_{\odot}$ each.
These large masses are also supported by the large (1300 km s$^{-1}$) 
velocity dispersion in 3C295 (Dressler \& Gunn 1992).
Both clusters may, therefore, have large Einstein angles of up to $\sim$ 1 arcmin, 
depending on the redshifts of the background sources (e.g. Kochanek 1992).  
One may naively expect an increased number of serendipitous sources near the 
clusters arising from the amplification of distant (and fainter) background sources.
Optical evidence for such an effect has been claimed by Rodrigues-Williams \& Hogan 
(1994) who report a statistically significant overdensity (by a factor 1.7) 
of high redshift QSOs in the directions of foreground galaxy clusters.
In 3C295, 9 (out of 17) X-ray sources are located in an annulus 
of radii $\sim$1$^{\prime}$-3$^{\prime}$ from the cluster 
center (see Fig. 1b), and 4 of these sources 
(sources n. 3, 5, 8 and 13, Table A.2) have redshifts larger than the 
cluster's redshift.

However, the magnification due to gravitational lensing produces two 
opposing effects on the source 
number counts (e.g. Croom \& Shanks 1999): 
Individual sources appear brighter by $\mu$ (where $\mu$ is the amplification 
factor) raising the number detected; but the area of sky lensed is reduced 
by the same factor $\mu$. The net effect depends on the slope of the 
logN-logS curve. 
At high fluxes where the logN-logS is steep, 
sources are added to counts; but at faint fluxes, the flatter logN-logS 
produces a {\it deficit} in the source counts (Wu 1994).   
Refregier \& Loeb (1997) predict on average a 
{\it reduction} in the surface density of faint resolved 
sources at fluxes fainter than about 10$^{-15}$ 
erg cm$^{-2}$ s$^{-1}$ (0.5-2 keV). 
An upturn in the logN-logS at very faint fluxes (e.g. $\lsimeq$ 2$\times$10$^{-16}$ erg 
cm$^{-2}$ s$^{-1}$) would be needed to create the observed excess.

\subsection{``Contamination'' of the $L_{\rm X}-T$ relation of high-z clusters}

Point sources contribute a substantial fraction of the 
clusters' fluxes, at least between 0.5-2 keV (see \S 3.2). 
Therefore, if this finding applies to a significant fraction of medium-high redshift clusters, 
past X-ray measurements of luminosities and temperatures of distant clusters 
obtained with low resolution experiments (see, e.g., the {\it ASCA} measurements 
by Mushotzky \& Scharf 1997) have probably been ``contaminated'' by 
the surrounding unresolved surplus sources (Note that the contribution from 
field sources is excluded through normal background subtraction). 
The present data indicate a possible contamination of the cluster's fluxes 
of up to $\sim$40\% and $\sim$15\% in RXJ0030 and 3C295, respectively. 
The effect on the observed $L_{\rm X}-T$ relation (e.g. Wu, Xue \& Fang, 1999) 
will depend on the way in 
which the contamination varies with redshift, with cluster size and on the surplus 
source spectra.
The sources have a summed spectrum of kT$\sim$5 keV (cluster rest-frame), therefore, 
at the 0$^{th}$-order, these would lower the observed temperature of clusters 
with kT$>$5 keV (i.e. the highest luminosity clusters) but increase 
that of lower-temperature clusters.
As a result, these effects should increase the scattering and possibly 
modify the shape of the true $L_{\rm X}-T$ relation of high-z clusters. 
Any changes to the $L_{\rm X}-T$ relation of clusters would have important
consequences for using clusters as cosmological probes (e.g., Henry 2000). 
More studies using $Chandra$ and $XMM$ are required in order to quantify 
the implications of such an effect.

\section{Conclusions}

We have presented a $Chandra$ study of an overdensity of X-ray sources 
around two z$\sim$0.5 clusters (RXJ0030 and 3C295). 
The observed source density near these clusters appears to be 
a factor of 2 times larger than expected on the basis of the 
{\it ROSAT} and {\it Chandra} logN-logS or by comparison to non-cluster $Chandra$
fields and to the outskirts of the cluster FOVs.
The radial distribution of the sources suggests that they are indeed 
concentrated within $\sim$200 arcsec of the clusters.
The effect is significant (at $\sim 2\sigma$/cluster or 3.5 $\sigma$ 
when combined) in the 0.5-2 keV energy band, but not in the 2-10 keV band
where the statistics are too poor. Deeper X-ray 
observations are needed to confirm unambiguously this result.
For both fields, the summed X-ray spectra of the faint sources 
are consistent with 
a power-law spectrum with $\Gamma \sim 1.7$ and no intrinsic 
absorption. If the sources are at the redshifts of the clusters, their average 
luminosity is $\sim$ 10$^{42-43}$ erg s$^{-1}$ making Seyfert-like 
AGNs/quasars the most likely counterparts. 
The number of redshifts currently available is, however, too small to be conclusive.

Possible explanations of the apparent overdensity may be:
a statistical variance of cosmic background sources; AGNs/quasars and/or powerful 
starburst galaxies associated with the clusters; and gravitationally lensed sources.
None of these explanations is without problems however. Only follow-up X-ray and 
optical observations (redshift measurements) will determine the true cause.
 

\acknowledgements
We wish to thank A. Dobrzycki, T. Aldcroft, P. Ciliegi, J. Drake, R. Della Ceca, 
G. Fabbiano, E. Falco, P. Freeman, F. La Franca, J. McDowell, S. Molendi, P. Nulsen, F. Tesch and 
V. Kashyap for many invaluable conversations and assistance. 
We are grateful to G.G.C. Palumbo and G. Zamorani for useful comments on an early version 
of the manuscript.
We especially acknowledge the entire $Chandra$ team which made 
these observations possible. M.C., P.M. and T.J.P. thank the Center for Astrophysics for hospitality.
M.C. was supported by NASA grants NAG5-3289 and NAG5-4808 while at CfA. 
M.C. and A.C. acknowledge partial support from ASI-ARS-98-119. P.M. acknowledges 
an ESA Research Fellowship. Some of us were also supported by the 
NASA contract NAS8-39073(CXC). E.J.H. was supported by NASA grant NAGW-3134.
This research has made use of the NASA/IPAC Extragalactic Database (NED) which 
is operated by the Jet Propultion Laboratory, California Institute of Technology, 
under contract with the National Aeronautics and Space Administration.


\centerline{APPENDIX}
\ss
\centerline{I - DATA CLEANING AND REDUCTION}
\ss

The data from the entire FOVs were cleaned and analyzed using the 
$Chandra$ Interactive Analysis of Observations (CIAO) software 
(release V1.1, Elvis et al. 2000, in preparation, see also http://asc.harvard.edu/cda/). 
The data were first filtered to include only the standard 
event grades 0, 2, 3, 4 and 6, and energies between 0.1--10 keV.
All hot pixels and bad columns were removed. Time intervals with
large background rate (i.e. larger than $\sim$ 3$\sigma$ over the quiescent 
value) were removed chip-by-chip yielding 
different exposures for each chip. The differences were small, 
$\lsimeq$ 5\%, for the 3C295, anti-Leonid and 3C273 fields but were $\sim$ 25\% for
RXJ0030, where the two BI chips in the RXJ0030 field had $\sim$ 
30 ks exposure compared to $\sim$ 40 ks for the FI chips.
This larger difference is due to the larger and more variable
background flux observed in the BI chips which, being
thinner, are less effective in rejecting background high-energy
particles (Markevitch 1999).
Data from the outer FI S4 chip and the BI S1 chip were excluded from the analysis 
of RXJ0030 and 3C295 because the effective exposure of S4 was 
$\sim$ 5 times smaller than for the other chips (indicative of a very 
noisy background) and because S1 was more than 10 arcmin 
from the optical axis, which reduces its sensitivity 
by 20-50\% due to vignetting.
For the anti-Leonid field, we also excluded the data from the S2 chip because of
its large ($>$ 10 arcmin) off-axis distance.
For the 3C273 field, we only considered the BI S3 chip, which is sufficient 
for the scope of the present study. We excluded in this field the area 
underneath the line emission which is due to exposure during read out.

The 3C273 field provides a comparison for the two cluster fields which is
independent of any systematic variations between ccds, since these 3 fields were all observed
with the same chip, S3. The larger area covered by the four FI chips in
the anti-Leonid observation lowers the statistical uncertainty in the non-cluster
number counts and allows a comparison of the sources spatial distribution (see \S4.3).
The FI chips have significantly lower background 
flux and a somewhat lower effective area at E$\lsimeq$1 keV than BI chips. 
However these effects should not significantly affect 
the present results because appropriate matrices 
(constructed with CIAO) were used to account for the different instrument 
responses and the background 
level is, in fact, negligible in both FI and BI at the flux limit used here (\S 4.1).


\ms
\centerline{II - THE SOURCE DETECTIONS}
\ss

In its simplest form, 
$wavdetect$ consists of correlating the image data with wavelet functions in
successively larger scale versions of the wavelet, comparing the
resulting ``correlation maps'' to a local background and 
detecting sources above a given threshold (Dobrzycki et al. 1999, Freeman et al. 2000). 
We considered only the 0.5-2 keV and 2-10 keV
energy bands where $Chandra$ is best calibrated, and which allows
direct comparison with {\it ROSAT}, {\it ASCA} and previous {\it Chandra} 
results (see \S4.1).
The original data were binned by 4, yielding pixels of $\sim$ 2'' on a side,
in order to obtain $\sim$ 1000$\times$1000 pixel images (for the whole FOV).
The $wavdetect$ software was run on several scales in 
order to match the (variable) dimensions of the PSF over the FOV.
Aspects of the detection method include: (a) the
computation of the correlation maps using a FFT; (b) the computation of a
local, exposure-corrected and normalized (i.e. flat-fielded) background in
each pixel; (c) its applicability to the low-counts regime of $Chandra$, 
as it does not
require a minimum number of background counts per pixel for accurate
computation of source detection thresholds; (d) its applicability to
multiscale data (i.e. extended sources); (e) its applicability to the 
$Chandra$ FOVs (i.e the algorithm recognises where the aim point is located); 
and (f) a full error analysis.
As a result, it is substantially more efficient than a sliding-cell 
technique in detecting weak point sources in 
crowded fields and extended sources.
 
We performed several runs of wavdetect, setting the probability of 
erroneously associating a background fluctuation in a pixel with a 
detection to 3.4 $\times$10$^{-6}$, 1.0$\times$10$^{-6}$,
2.9$\times$10$^{-7}$, and 1.0$\times$10$^{-7}$. 
These probabilities correspond 
approximately to Gaussian equivalent significances of 4.5, 4.7, 5 
and 5.2$\sigma$, respectively. ACIS-I simulations and other 
calculations (Freeman et al. 2000) show that these thresholds produce 
$\leq$ 3, 1, 0.3 and 0.1 false detections per 4 chips FOV, respectively.
We set the threshold to correspond to one expected spurious source 
(probability of 10$^{-6}$, 4.7$\sigma$).

The $Chandra$ dither pattern creates rapid changes in exposure 
near the edges of the chips which are not currently accounted for in either 
source detection algorithm. As a result, this introduces a significant 
number of spurious detections of extremely faint sources with 
$\lsimeq$ 6 counts. We therefore introduced an additional selection 
criterion: that the sources should also have a significance 
level SN$>$3, where SN (an output of $wavdetect$) is defined 
as the number of source counts divided by the Gehrels (1986) standard 
deviation of the number of background counts, in order to be 
considered as real.
We checked and found that none of the excluded sources had a flux larger 
than the flux limit used in \S 4.1.

We also checked the results from $wavdetect$ 
by comparing them to the $celldetect$ (Dobrzycki et al. 1999) output.
$celldetect$ is a robust sliding-cell algorithm in CIAO
which utilizes a sliding-box of variable size according 
to the position in the FOV and according to a pre-defined encircled 
energy (for a PSF calculated at a given energy).  
The background is calculated, locally, in annuli centered 
on the cells with an area equal to the source area 
(typical sizes of the boxes were 6--10'' per side, on-axis, 
and 10-15'' per side, at $\lsimeq$ 5$^\prime$ off-axis). 
We found that, for sources brighter than $\sim$ 15 net counts, the match was better 
than $\sim$ 90\% in both the obtained positions and number of source 
counts, confirming that the wavelet algorithm is working well.

\ms
\centerline{III - THE SOURCE COUNTS AND FLUXES}
\ss

Source count rates were obtained using $wavdetect$ from regions with 
typical radii of $\sim$ 3'' (on-axis) and 10'' (off-axis).
Comparing the $wavdetect$ cell regions with the expected dimensions of
the PSF at the source positions, we estimate that our extraction
regions contain more than 99\% of the PSF encircled energy.
Background counts were estimated locally for the same area on the 
basis of the background map produced by $wavdetect$ 
(Freeman et al. 2000). 

The measured counts were first corrected for vignetting
and then converted to an emitted, unabsorbed flux. 
The vignetting is larger for photons with E$>$ 4 keV and, at those energies, 
the effective area is reduced by $\sim$ 40\% at 10 arcmin. 
We applied a correction for the effective area off-axis using the 
pre-flight calibration. 
However, this correction was typically not large, since our detections
are dominated by photons with E $< 4$ keV (see Fig. 5). The vignetting 10$'$ off axis
drops to $\sim 15\%$ for photon energies of $\sim 1.5$ keV, where Chandra's
effective area peaks.
The correction we applied is based on Fig. 4.3 of the 
``{\it Chandra} Proposer's Observatory Guide'' (1999) 
that we approximate using a linear function of the form: 
$$ V_{\rm cor} = 0.97 + 0.0175 \times \theta_{\rm off}$$ 
where $V_{\rm cor}$ is the vignetting correction factor (to be multiplied to all the
counts) and $\theta_{\rm off}$ is the off-axis distance in arcmin.

For the conversion from counts to flux, a power-law 
spectrum with photon index $\Gamma$=2 and $\Gamma$=1.7 
was assumed for the sources detected between 0.5-2 keV and 2-10 keV, respectively.
We assumed Galactic absorption columns of 3.91, 1.33, 2.45 and 1.8 $\times$
10$^{20}$ cm$^{-2}$ along the line of sight of the RXJ0030, 3C295, anti-Leonid and 3C273 
fields, respectively (Stark et al. 1992).
These spectral models were chosen to allow a direct comparison
with the {\it ROSAT} PSPC and {\it ASCA} results (Hasinger et al. 1998, Della Ceca et al. 2000). 
These assumed spectra are consistent with the summed
spectra obtained in \S 4.2. Moreover in these
energy bands, the fluxes are only weakly dependent
on the spectral slope and the Galactic $N_{\rm H}$ value adopted (e.g. a $\Delta
\Gamma$=0.2 gives a $\Delta$flux of less than 5\%).
Systematic instrumental errors in the absolute
effective area and energy scale are conservatively below
10\%-20\% at all the energies considered here 
(N. Schultz, private communication). Small changes can be expected once 
refined calibrations become available.

\vfill\eject

\begin{table}[htb]
\begin{center}
\vspace{-0.5cm}
{\bf Table 1} Log of Observations\\
\vspace{0.2cm}
\begin{tabular}{cccccc}
\hline
\hline
\multicolumn{1}{c}{} &
\multicolumn{1}{c}{Obs. Id} &
\multicolumn{1}{c}{Obs. Mode} &
\multicolumn{1}{c}{Exposure$^*$} &
\multicolumn{2}{c}{Nominal Pointing} \\
\multicolumn{1}{c}{} &
\multicolumn{1}{c}{} &
\multicolumn{1}{c}{} &
\multicolumn{1}{c}{(s)} &
\multicolumn{1}{c}{RA(2000)} &
\multicolumn{1}{c}{DEC(2000)} \\
\hline
\multicolumn{6}{c}{\bf Cluster Fields} \\
RXJ0030 & 1190 & ACIS-S, 4 chips & 15375 & 00:30:40 & +26:18:00 \\
        & 1226 & ACIS-S, 4 chips & 14746 &            &         \\
        & merged& ACIS-S, 4 chips & 30121&            &          \\
3C295   &  578 & ACIS-S, 4 chips & 18280 & 14:11:10 & +52:13:01 \\
\multicolumn{6}{c}{\bf Comparison Fields} \\
anti-Leonid  & 1479 & ACIS-I, 4 chips & 20580 & 22:13:12 & $-$22:10:41 \\
3C273   & 1712 & ACIS-S, 1 chip(S3) & 22800 & 12:29:06 & +02:03:14 \\
\hline
\hline
\end{tabular}
\end{center}
\normalsize
$^*$ Exposure time obtained for the on-axis chips (ACIS-S3 for RXJ0030, 3C295 and
3C273, and ACIS-I3 for anti-Leonid)
\end{table}

\begin{table}[htb]
\vspace{-0.5cm}
\begin{center}
{\bf Table 2a} Number of Sources Detected between 0.5-2 keV \\
\vspace{0.3cm}
\begin{tabular}{cccccc}
\hline
\hline
\multicolumn{1}{c}{} &
\multicolumn{1}{c}{RXJ0030(S3)} &
\multicolumn{1}{c}{3C295(S3)} &
\multicolumn{1}{c}{{\it ROSAT}$^a$/{\it Chandra}$^b$/{\it Chandra}$^c$} &
\multicolumn{1}{c}{anti-Leonid(I0-3)$^d$} &
\multicolumn{1}{c}{3C273(S3)} \\
\hline
\multicolumn{6}{c}{F$_{limit}$=3$\times$ 10$^{-15}$ erg~cm$^{-2}$~s$^{-1}$} \\
\hline
$CHIP^{-1}$ & 13 $\pm$ 3.6 & 13 $\pm$ 3.6 &             & 7.7 $\pm$ 1.4 & 5 $\pm$ 2.2   \\
$Deg^{-2}$ & 731 $\pm$ 202 & 731 $\pm$ 202 & 336$\pm$31/335/288 & 436$\pm$78    & 280 $\pm$ 126\\
\hline
\multicolumn{6}{c}{F$_{limit}$ = 1.5 $\times$ 10$^{-15}$ erg~cm$^{-2}$~s$^{-1}$} \\
\hline
$CHIP^{-1}$ & 21 $\pm$ 4.6 & 16 $\pm$ 4 & & 10.5$\pm$1.6  & 11 $\pm$ 3.3\\
$Deg^{-2}$ & 1181 $\pm$ 259 & 900$\pm$225 & 600$\pm$70/544/502 & 578$\pm$64 & 619 $\pm$ 187\\
\hline
\hline
\end{tabular}
\end{center}
\begin{center}
{\bf Table 2b} Number of Sources Detected between 2-10 keV \\
\vspace{0.3cm}
\begin{tabular}{cccccc}
\hline
\hline
\multicolumn{1}{c}{} &
\multicolumn{1}{c}{RXJ0030(S3)} &
\multicolumn{1}{c}{3C295(S3)} &
\multicolumn{1}{c}{{\it ASCA}/model$^e$} &
\multicolumn{1}{c}{anti-Leonid(I0-3)$^d$} &
\multicolumn{1}{c}{3C273(S3)} \\
\hline
\multicolumn{6}{c}{F$_{limit}$ = 2 $\times$ 10$^{-14}$ erg~cm$^{-2}$~s$^{-1}$} \\
\hline
$CHIP^{-1}$ & 4 $\pm$ 2 & 4 $\pm$ 2 &  & 1.2$\pm$0.5 & 4 $\pm$ 2\\
$Deg^{-2}$ & 225 $\pm$ 112 & 225 $\pm$ 112  & 120/120 & 70$\pm$31 & 225 $\pm$ 112\\
\hline
\hline
\end{tabular}
\end{center}
\normalsize
$^a$ {\it ROSAT} PSPC values and statistical errors obtained from Hasinger et al. (1998)

$^b$ {\it Chandra} ``best-fit'' values obtained from Mushotzky et al. (2000)

$^c$ {\it Chandra} ``best-fit'' values obtained from Giacconi et al. (2000)

$^d$ sources were detected in the 4 ACIS-I chips and normalized 
for 1 single chip (for comparison with RXJ0030, 3C295 and 3C273).

$^e$ Linear extrapolation of measured {\it ASCA} values from Della Ceca et al. (2000) 
and expected value from theoretical model of Comastri et al. (1999).
\pn
NOTE: Errors on the $Chandra$ numbers are the square root of N.
\end{table}

\begin{table}[htb]
\begin{center}
\vspace{-0.5cm}
{\bf Table 3:} Best-fits of summed$^*$ spectra - Single power-law models \\
\begin{tabular}{ccccccccc}
\hline
\hline
Field & $N_{\rm H}\equiv N_{\rm Hgal}$ & $\Gamma$ & $\chi^{2} (d.o.f)$\\
      & ($\times 10^{20}$ cm$^{-2}$) & &  \\ 
\hline
RXJ0030   & 3.9 & 1.72$^{+0.13}_{-0.12}$ &  27.5 (39) \\
3C295 & 1.33 & 1.79$^{+0.15}_{-0.13}$ & 21.3 (30) \\
anti-Leonid & 2.45 & 2.28$^{+0.17}_{-0.15}$ & 34.7 (42)\\
3C273 & 1.80 & 1.24$^{+0.25}_{-0.20}$ & 32.9 (36) \\
\hline
\hline
\end{tabular}
\end{center}
$^*$ Computed with 22 sources in RXJ0030, 16 in 3C295, 24 in the
anti-Leonid fields, and 12 in 3C273 (see \S 5.1).

Note: Intervals are at 90\% confidence for one interesting parameter.
\end{table}

\begin{figure}[htb]
{\psfig{file=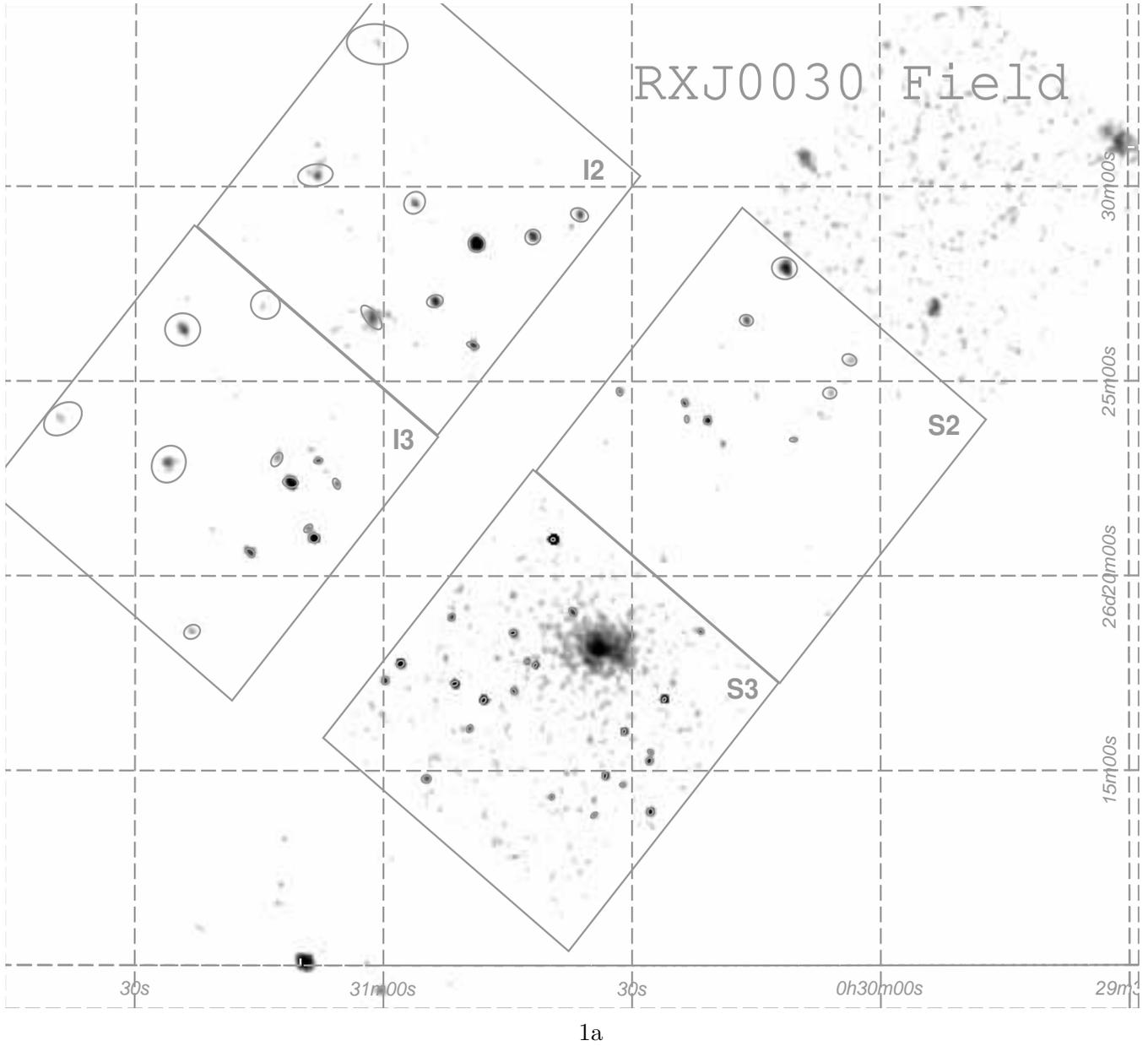,width=18cm,height=16cm,angle=0}}
\centerline{1a}
\caption[h] {Images of the (a) RXJ0030, (b) 3C295, (c) anti-Leonid and (d) 3C273 fields
between 0.5-2 keV. The images (north up and east right) were binned by a factor of 4 
(i.e. 1 pixel$\simeq$2 arcsec) and gaussian-smoothed with a $\sigma$=1 pixel. 
The sources detected as serendipitous sources (those listed in Table A) are marked with 
elliptical regions. Their sizes are set by deriving the 1$\sigma$
principal axes (and rotation angle) of the source counts distribution
for each source, and multiplying these by 10 for greater visual
clarity. Chip boundaries are marked by the line. Only the chips outlined 
were used in the analysis (see App. I). The line of emission in the 3C273 field 
is due to exposure during read out which is known to occur with very bright sources.}
\end{figure}
\begin{figure}
{\psfig{file=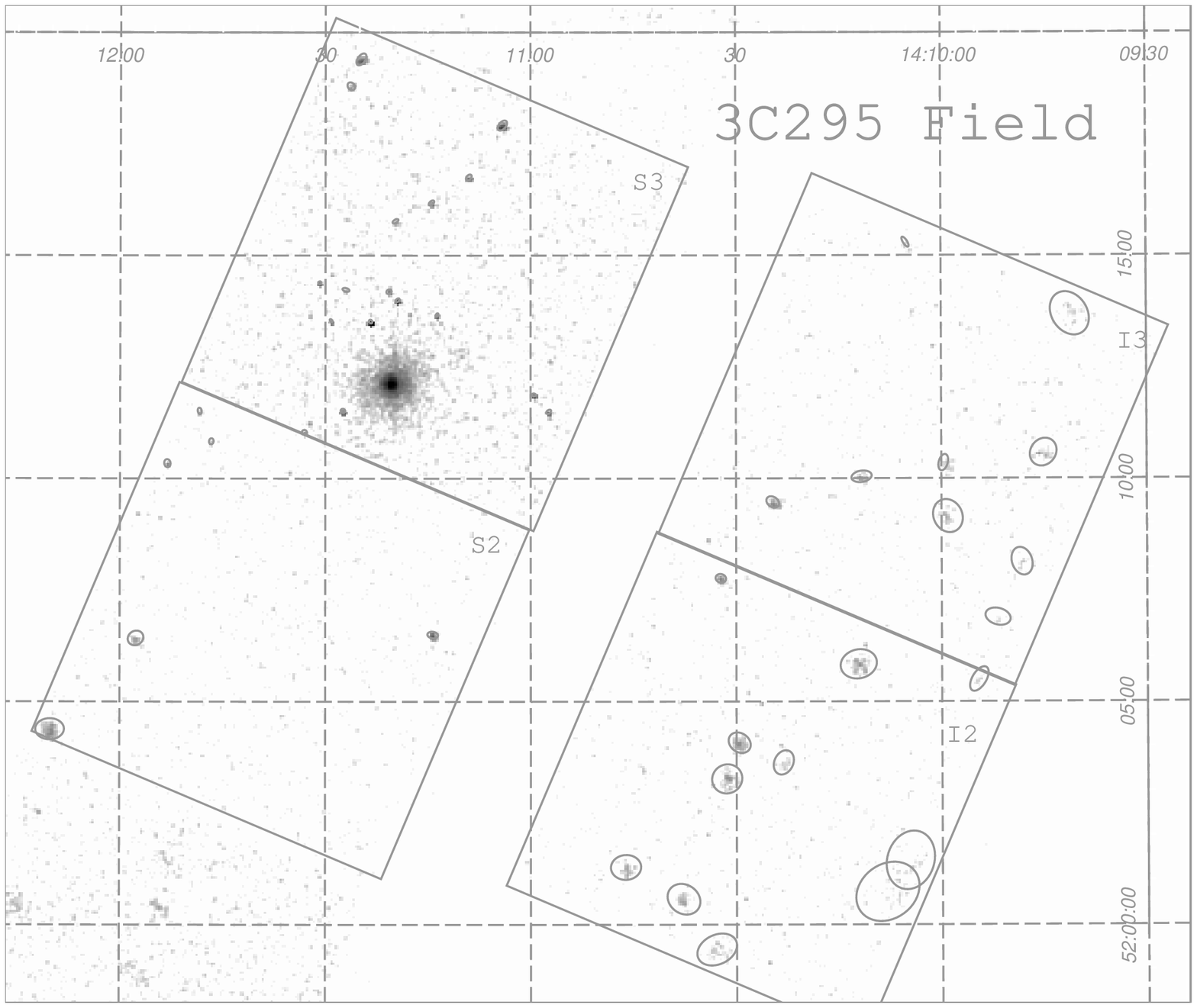,width=18cm,height=16cm,angle=0}}
\centerline{1b}
\end{figure}
\begin{figure}
{\psfig{file=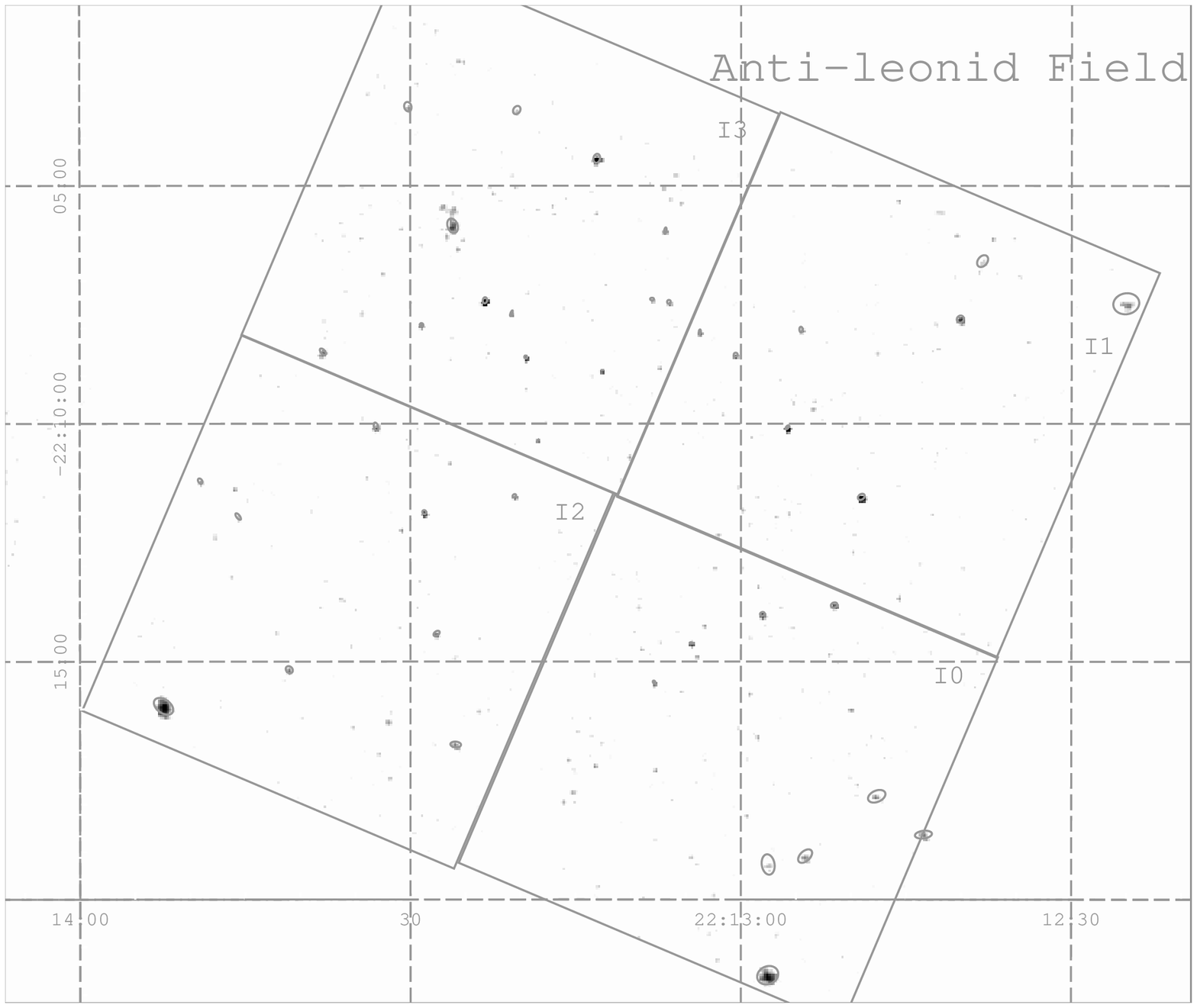,width=18cm,height=16cm,angle=0}}
\centerline{1c}
\end{figure}
\begin{figure}
{\psfig{file=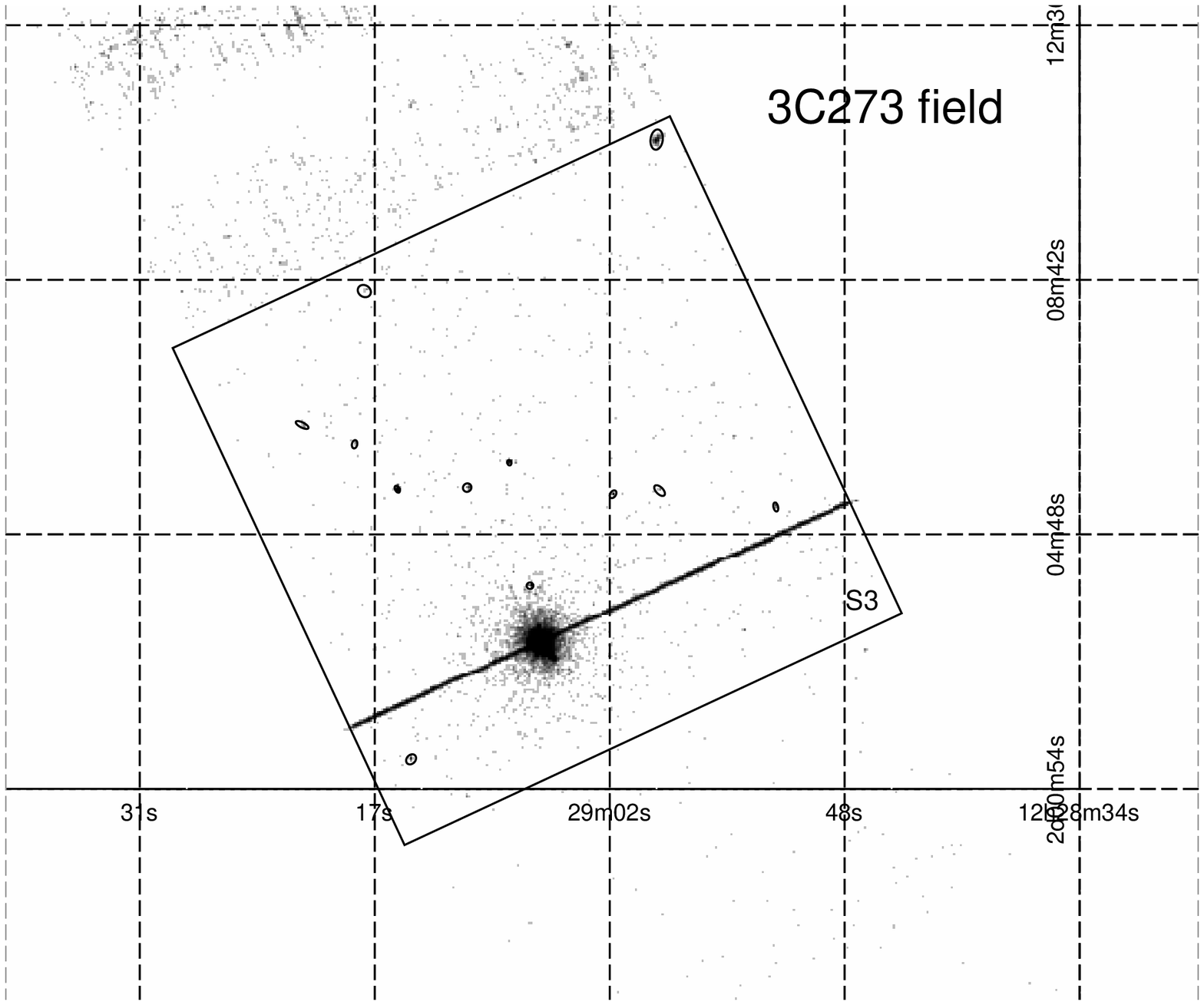,width=18cm,height=16cm,angle=0}}
\centerline{1d}
\end{figure}

\begin{figure}[htb]
\parbox{15truecm}
{\psfig{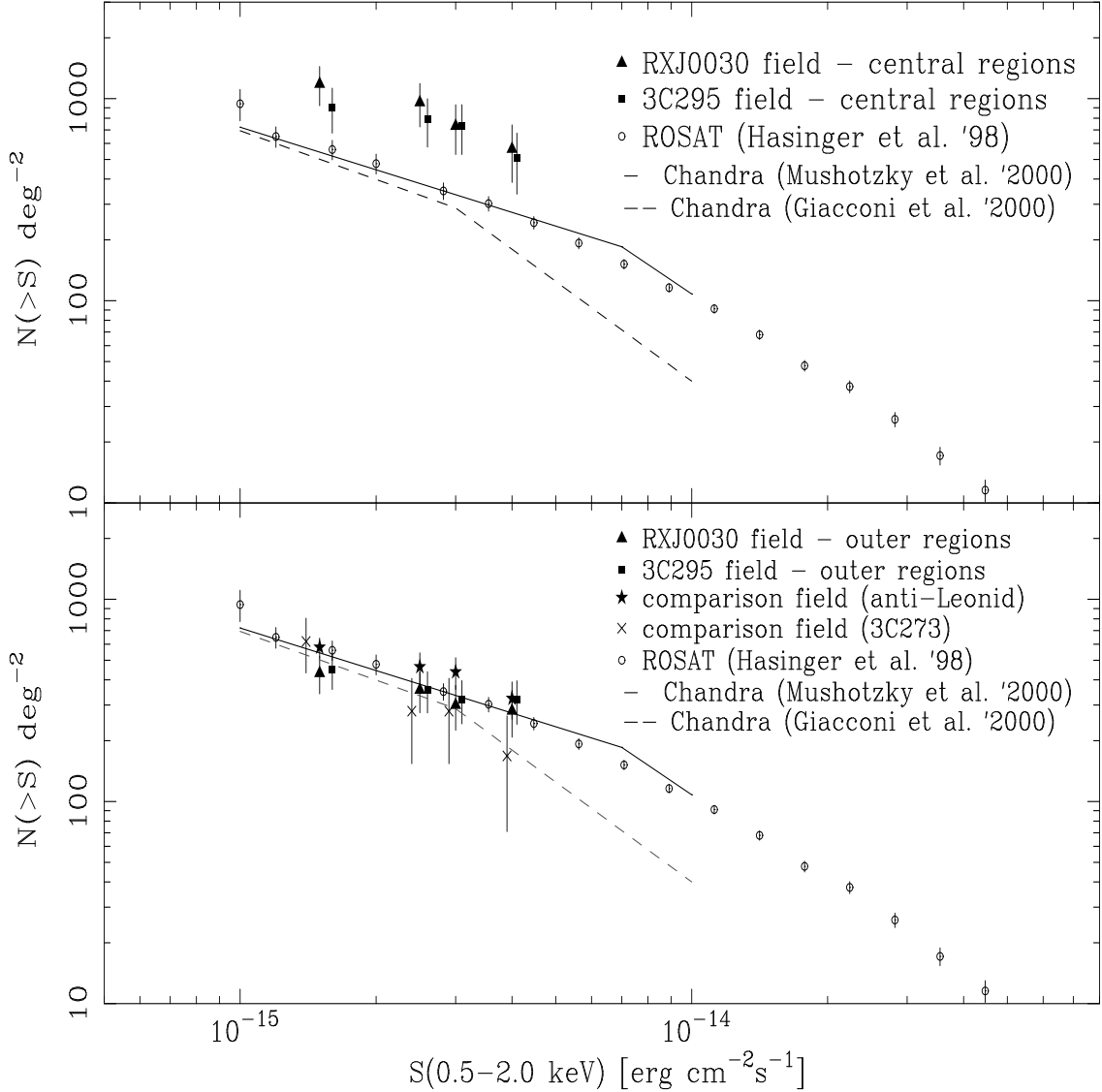}}
\par\noindent
\caption[h] {{\it Upper panel}: $Chandra$ LogN-logS between 0.5-2 keV measured in 
the RXJ0030 and 3C295 fields (only central chips, see text for details) compared 
to the 0.5-2 keV logN-logS measured by {\it ROSAT} (PSPC+HRI) from Hasinger et al. (1998) and 
$Chandra$ ``best-fit'' results from Mushotzky et al. (2000) and Giacconi et al. (2000).
In the Giacconi et al. logN-logS, the lack of sources for fluxes between 
3$\times$10$^{-15}$ and 10$^{-14}$ erg s$^{-1}$ cm$^{-2}$ is attributed to cosmic 
variance (Giacconi et al. 2000).
Reported fluxes for 3C295 were increased by 1$\times$10$^{-16}$ erg cm$^{-2}$ s$^{-1}$ for plotting 
clarity. {\it Lower panel}: $Chandra$ LogN-logS between 0.5-2 keV measured in 
the RXJ0030 and 3C295 fields (only outer chips) and logN-logS obtained 
in the two comparison fields (Anti-Leonid: whole FOV; 3C273: chip S3). 
{\it ROSAT} and $Chandra$ deep-field data points are again plotted for comparison. 
Reported fluxes for 3C273 were 
decreased by 10$^{-16}$ erg cm$^{-2}$ s$^{-1}$ for plotting clarity.}
\end{figure}

\begin{figure}[htb]
\parbox{14truecm}
{\psfig{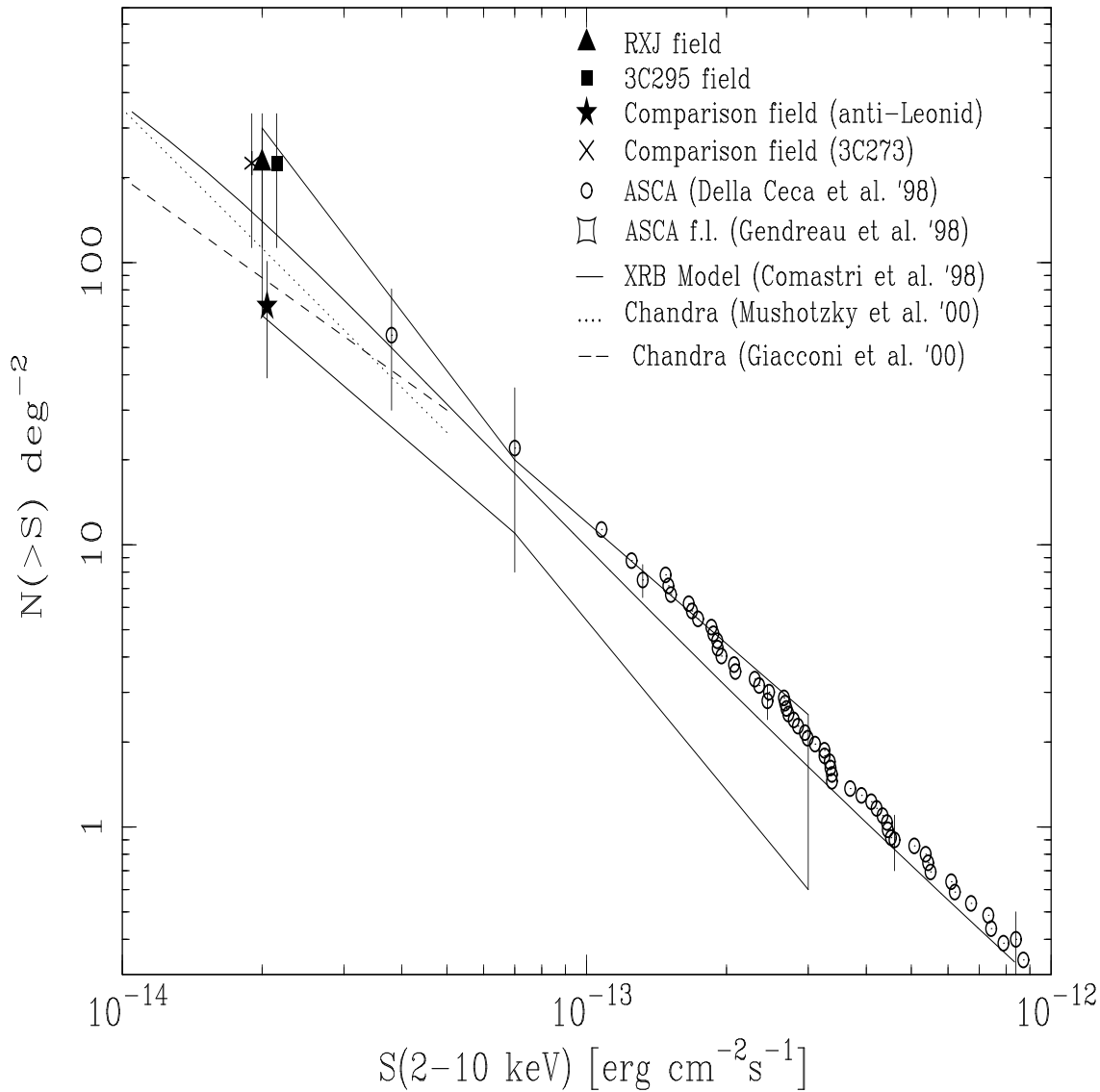}}
\par\noindent
\caption[h] {$Chandra$ LogN-logS between 2-10 keV measured for a flux limit 
of 2$\times$10$^{-14}$ erg cm$^{-2}$ s$^{-1}$ in the RXJ0030, 3C295 
(only central chips, see text for details), 3C273 (S3 chip) and anti-Leonid 
fields (whole FOV, 4 chips). 
{\it ASCA} points (Cagnoni et al. 1998, Della Ceca et al. 2000), {\it ASCA} fluctuation limits 
(Gendreau et al. 1998), an AGN model for the XRB (Comastri et al. 1995) 
and {\it $Chandra$ deep-field ``best-fit'' results 
(calculated between 1--5$\times$10$^{-14}$ erg cm$^{-2}$ s$^{-1}$ from Mushotzky et al. 2000, 
Giacconi et al. 2000) are plotted for comparison.}}
\end{figure}


\vfill\eject

\begin{figure}[htb]
\parbox{6truecm}
{\psfig{file=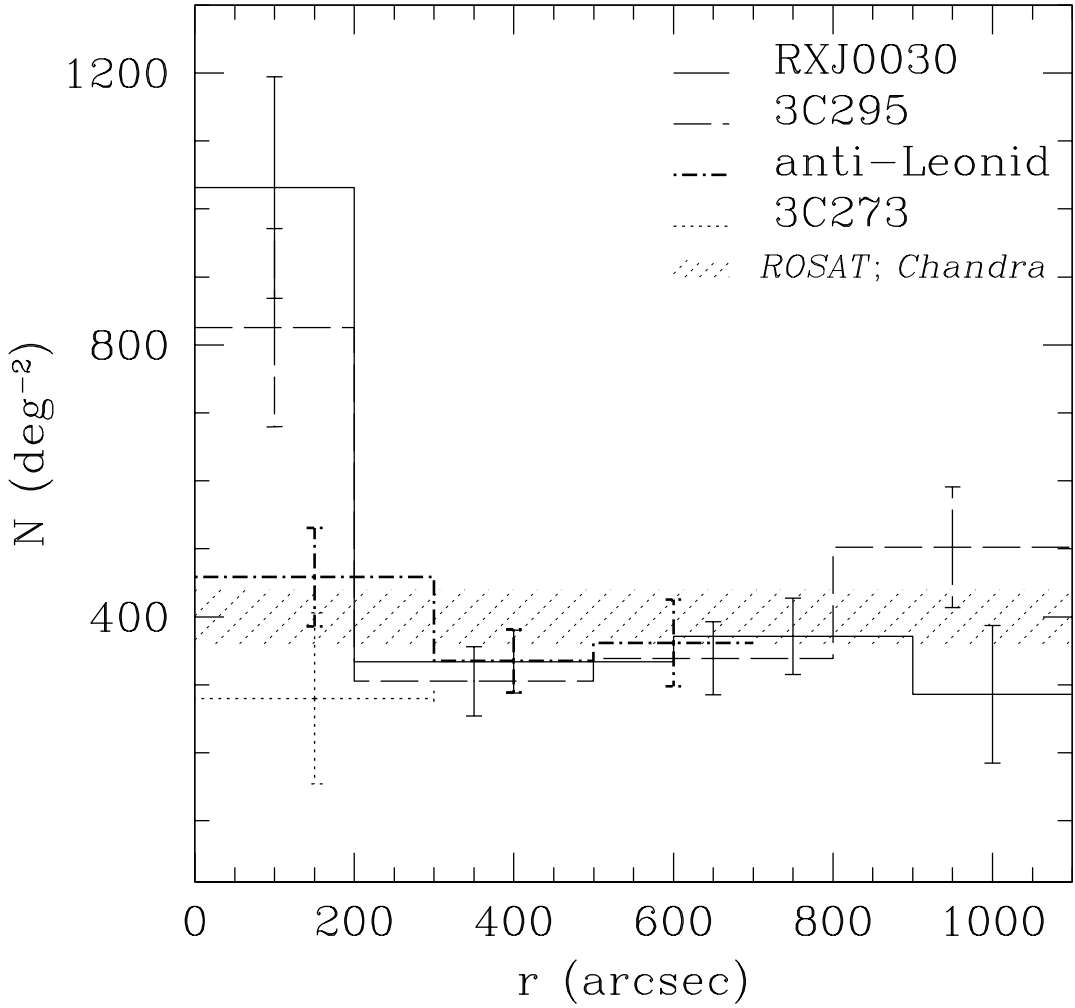,width=15cm,height=15cm,angle=0}}
\par\noindent
\caption[h]{Radial distribution of the sources in RXJ0030 (solid line), 
3C295 (dashed line), anti-Leonid (thick dash-dot line), and 3C273 (dotted line). 
At z=0.5, 200 arcsec 
correspond to $\sim$ 1.4 Mpc. The dashed area corresponds to the level of the 
{\it ROSAT} and {\it Chandra} logN-logS at a flux limit of 
2.5$\times$10$^{-15}$ erg cm$^{-2}$ s$^{-1}$.}

\end{figure}

\vfill\eject

\begin{figure}[htb]
\parbox{6truecm}
{\psfig{file=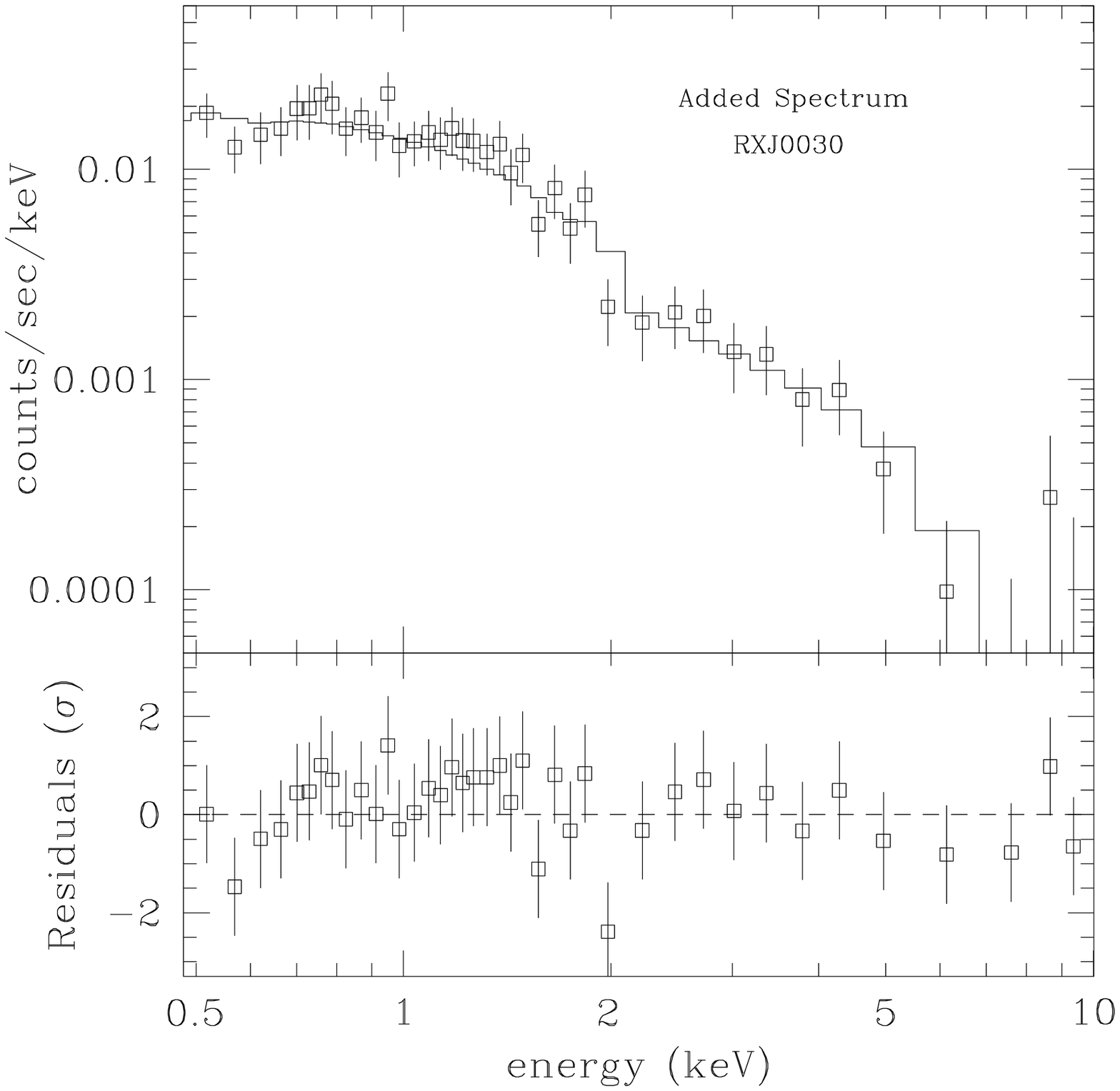,width=14cm,height=6cm,angle=0}}
\ \hspace{0.5truecm} \
\parbox{6truecm}
{\psfig{file=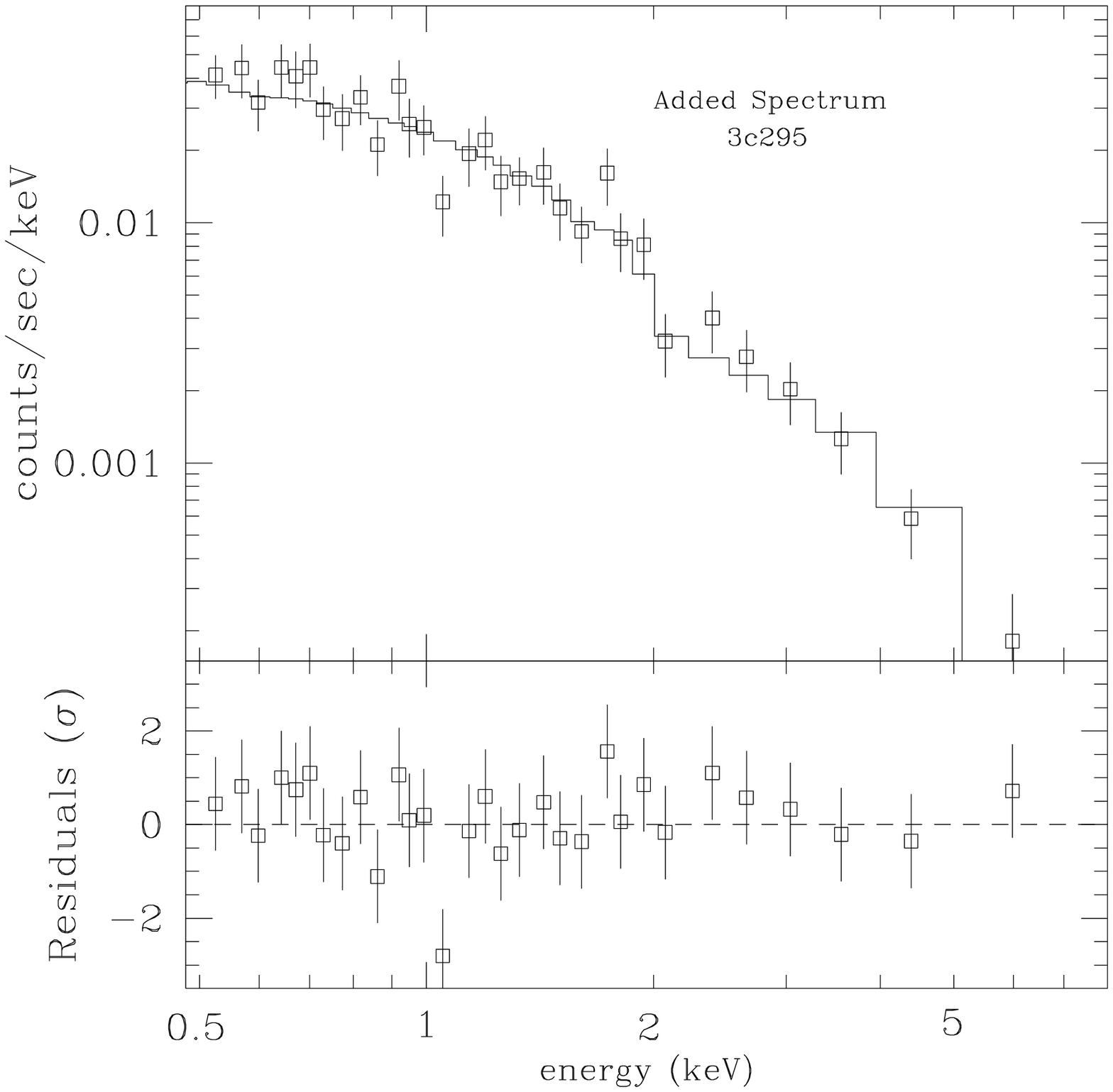,width=14cm,height=6cm,angle=0}}
\ \hspace{1.5truecm} \
\parbox{6truecm}
\ \hspace{0.5truecm} \
{\psfig{file=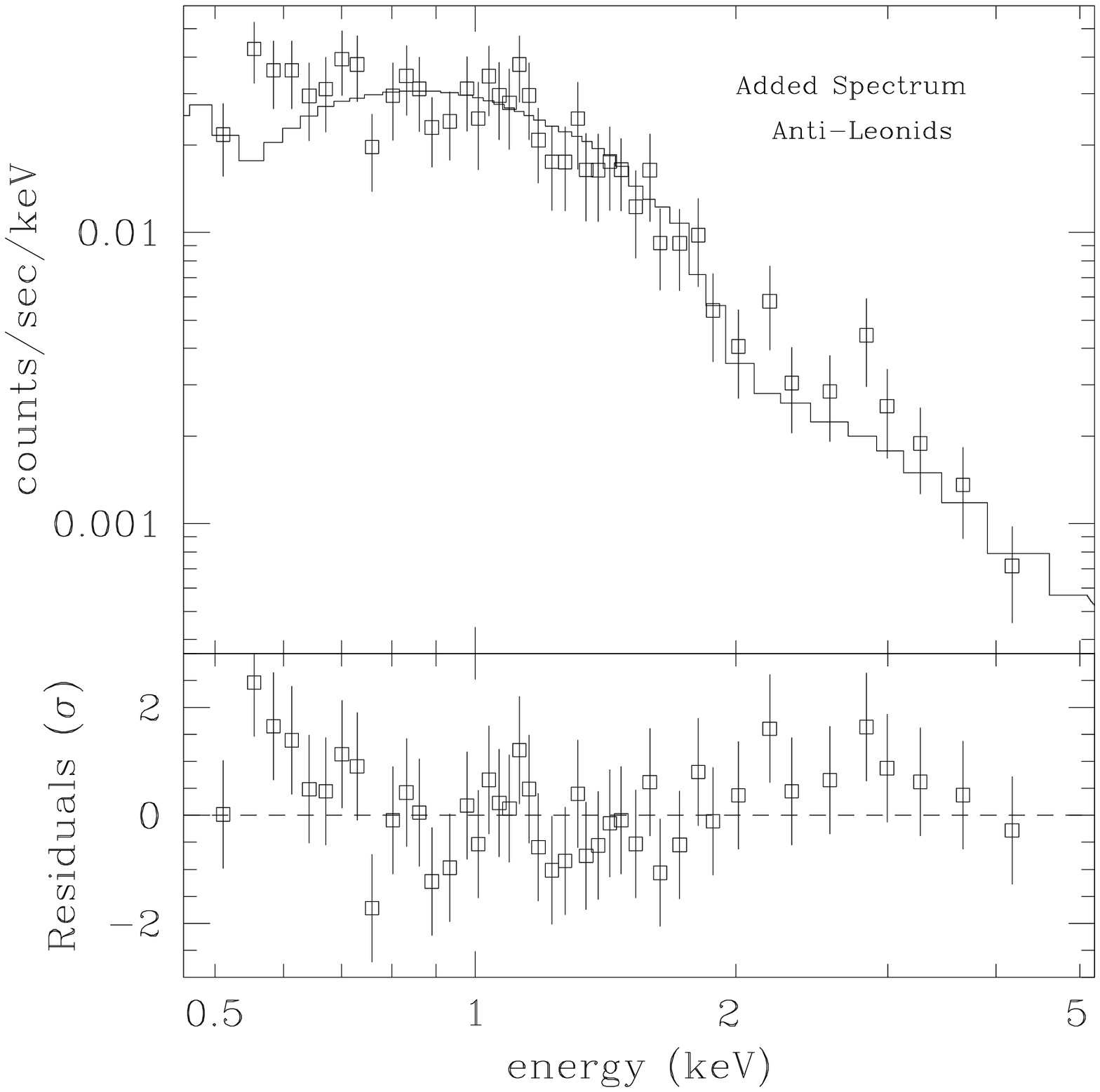,width=14cm,height=6cm,angle=0}}
\parbox{6truecm}
{\psfig{file=./3c273_spec_bw.ps,width=12cm,height=6cm,angle=-90}}
\parbox{6truecm}
\par\noindent
\caption[h] {Added spectra of all the serendipitous sources (cluster and/or brightest 
sources excluded) in the RXJ0030, 3C295, anti-Leonid, and 3C273 fields: best-fit spectra 
modelled with a single power-law as given in Table 3. Data were binned in order to have a S/N$>$3 in each energy bin.}
\end{figure}

\vfill\eject

\hspace{-0.5cm}{
\begin{table}
\small
\begin{center}
{\bf Table A - 1:} list of point sources detected between 0.5-2 keV and 2-10 keV \\
in the RXJ0030 field -- ACIS-S3 central chip (target cluster excluded)
\tiny
\begin{tabular}{cccccccccccccc}
\hline
\hline
\multicolumn{14}{c}{\bf \small RXJ0030 field} \\
\hline
\multicolumn{1}{c}{row} &
\multicolumn{1}{c}{ra$^a$ (2000)} &
\multicolumn{1}{c}{dec$^a$ (2000)} &
\multicolumn{1}{c}{Net Counts} &
\multicolumn{1}{c}{Bgd Counts} &
\multicolumn{1}{c}{Flux} &
\multicolumn{1}{c}{$\theta_{\rm off}$} &
\multicolumn{1}{c}{$B$} &
\multicolumn{1}{c}{$\alpha_{\rm ox}^b$} &
\multicolumn{1}{c}{z} &
\multicolumn{1}{c}{$L_{\rm X}^c$} &
\multicolumn{1}{c}{$M_{B}$} &
\multicolumn{1}{c}{class} &
\multicolumn{1}{c}{Ref.$^d$} \\
\multicolumn{1}{c}{} &
\multicolumn{1}{c}{} &
\multicolumn{1}{c}{} &
\multicolumn{1}{c}{} &
\multicolumn{1}{c}{} &
\multicolumn{1}{c}{10$^{-15}$cgs} &
\multicolumn{1}{c}{($^\prime$)} &
\multicolumn{1}{c}{} &
\multicolumn{1}{c}{} &
\multicolumn{1}{c}{} &
\multicolumn{1}{c}{10$^{42}$erg/s} &
\multicolumn{1}{c}{} &
\multicolumn{1}{c}{} &
\multicolumn{1}{c}{} \\
\\
\hline
\hline
\multicolumn{14}{c}{\bf \small (0.5-2 keV)} \\
 1 &  00:30:39.5 & +26:20:56.4 & 1023.49 $\pm$ 32.47 & 6.76 $\pm$ 0.03 & 112.58 $\pm$ 3.57 & 3.0 & 17.3 & 1.62 & 0.492 & 134.2 & -25.2 & QSO & 1 \\ 
 2 &  00:30:57.9 & +26:17:44.8 & 120.54 $\pm$ 11.49 & 4.01 $\pm$ 0.02 & 13.26 $\pm$ 1.26 & 5.0 & & & & - & & & \\ 
 3 &  00:30:47.9 & +26:16:48.6 & 96.23 $\pm$ 10.17 & 4.85 $\pm$ 0.02 & 10.59 $\pm$ 1.12 & 3.0 & 18.4 & 1.85 & 0.246 & 2.9 & -22.6 & Sey2 & 2\\ 
 4 &  00:30:51.4 & +26:17:14.3 & 84.14 $\pm$ 9.63 & 5.54 $\pm$ 0.03 & 9.26 $\pm$ 1.06 & 3.6 & 19.7 & 1.66 & 0.129 & 0.7 &  -19.6 & gal? & 3\\ 
 5 &  00:30:26.1 & +26:16:49.7 & 83.06 $\pm$ 9.36 & 3.20 $\pm$ 0.02 & 9.14 $\pm$ 1.03 & 2.6 & 19.2 &  1.76 & 0.269& 3.0 & -21.9 & QSO & 3\\ 
 6 &  00:30:27.8 & +26:13:60.0 & 49.97 $\pm$ 7.50 & 3.41 $\pm$ 0.02 & 5.50 $\pm$ 0.82 & 4.6 & 19.8 & 1.79 & & - & & &\\ 
 7 &  00:30:27.9 & +26:15:15.2 & 38.33 $\pm$ 6.67 & 4.77 $\pm$ 0.02 & 4.22 $\pm$ 0.73 & 3.4 & & & & -  & & &\\ 
 8 &  00:30:41.7 & +26:17:42.7 & 38.04 $\pm$ 6.44 & 3.69 $\pm$ 0.02 & 4.18 $\pm$ 0.71 & 1.3 & & & & - & & &\\ 
 9 &  00:30:33.3 & +26:14:52.4 & 37.59 $\pm$ 6.50 & 3.42 $\pm$ 0.02 & 4.14 $\pm$ 0.71 & 3.3 & 19.3 & 1.86 & 0.247 & 1.1 & -21.3 & gal. & 3\\ 
 10 &  00:30:44.4& +26:18:32.6 & 37.45 $\pm$ 6.43 & 3.71 $\pm$ 0.02 & 4.12 $\pm$ 0.71 & 2.0 & & & & - & & &\\ 
 11 &  00:30:31.0& +26:16:00.6 & 35.43 $\pm$ 6.24 & 2.98 $\pm$ 0.02 & 3.90 $\pm$ 0.69 & 2.4 & & & & - & & &\\ 
 12 &  00:30:59.8& +26:17:18.9 & 34.80 $\pm$ 6.30 & 2.34 $\pm$ 0.02 & 3.83 $\pm$ 0.69 & 5.5 & & & & - & & &\\ 
 13 &  00:30:54.8& +26:14:47.5 & 28.81 $\pm$ 6.02 & 4.94 $\pm$ 0.02 & 3.17 $\pm$ 0.66 & 5.4 & & & & - & & &\\ 
 14 &  00:30:37.2& +26:19:04.4 & 25.79 $\pm$ 5.68 & 6.91 $\pm$ 0.03 & 2.84 $\pm$ 0.62 & 1.1 & & & & - & & &\\ 
 15 &  00:30:51.8& +26:18:56.6 & 25.34 $\pm$ 5.47 & 3.51 $\pm$ 0.02 & 2.79 $\pm$ 0.60 & 3.7 & & & & - & & &\\ 
 16 &  00:30:21.7& +26:18:34.4 & 23.93 $\pm$ 5.24 & 2.72 $\pm$ 0.02 & 2.63 $\pm$ 0.58 & 3.3 & & & & - & & &\\ 
 17 &  00:30:44.3& +26:17:02.7 & 23.62 $\pm$ 5.23 & 3.56 $\pm$ 0.02 & 2.60 $\pm$ 0.58 & 2.1 & 20.2 & 1.81 &  & -  & & &\\ 
 18 &  00:30:49.6& +26:16:05.1 & 19.93 $\pm$ 4.96 & 3.73 $\pm$ 0.02 & 2.19 $\pm$ 0.55 & 3.7 & & & & - &  & &\\ 
 19 &  00:30:42.7& +26:17:48.6 & 19.60 $\pm$ 4.68 & 2.33 $\pm$ 0.02 & 2.16 $\pm$ 0.51 & 1.5 & & & & - &  & &\\ 
 20 &  00:30:39.7& +26:14:19.5 & 18.15 $\pm$ 4.64 & 2.51 $\pm$ 0.02 & 2.00 $\pm$ 0.51 & 3.9 & 19.1 & 2.03 & & - & & &\\ 
 21 &  00:30:27.8& +26:15:28.9 & 14.61 $\pm$ 4.23 & 2.76 $\pm$ 0.02 & 1.61 $\pm$ 0.47 & 3.2 & & & & - &  & &\\ 
 22 &  00:30:34.6& +26:13:51.9 & 13.46 $\pm$ 4.18 & 3.11 $\pm$ 0.02 & 1.48 $\pm$ 0.46 & 4.3 & & & & - &  & &\\ 
 23 &  00:30:31.2& +26:14:38.3 & 11.84 $\pm$ 3.87 & 2.54 $\pm$ 0.02 & 1.30 $\pm$ 0.43 & 3.6 & & & & - &  & &\\ 
\multicolumn{14}{c}{\bf \small (2-10 keV)}\\
 1 &  00:30:39.5 & +26:20:56.2 & 596.50 $\pm$ 25.10 & 19.01 $\pm$ 0.05 & 417.55 $\pm$ 17.57 & & & & & 497.7 &  & &\\ 
 4 &  00:30:51.4 & +26:17:14.8 & 37.38 $\pm$ 7.23 & 12.81 $\pm$ 0.04 & 26.16 $\pm$ 5.06 & & & & & 34.4 & & &\\ 
 3 &  00:30:47.9 & +26:16:48.9  &  32.51 $\pm$ 6.86 & 13.19 $\pm$ 0.04 & 22.76 $\pm$ 4.80 & & &  & & 6.3 & & &\\ 
 11 &  00:30:31.0 & +26:16:00.5  & 28.10 $\pm$ 6.40 & 12.23 $\pm$ 0.04 & 19.67 $\pm$ 4.48 & & & & & - & & &\\ 
  2 &  00:30:57.8 & +26:17:44.8 & 16.44 $\pm$ 4.96 & 6.45 $\pm$ 0.03 & 11.51 $\pm$ 3.47 & & & & & - & & &\\ 
 5 &  00:30:26.2 & +26:16:49.6 & 14.60 $\pm$ 4.65 & 6.61 $\pm$ 0.03 & 10.22 $\pm$ 3.26 & & & & & 13.5 & & &\\ 
\hline
\end{tabular}
\end{center}
\end{table}

\vfill\eject

\begin{table}[htb]
\begin{center}
\small
{\bf Table A - 2:} list of point sources detected between 0.5-2 keV and 2-10 keV \\
in the 3C295 field -- ACIS-S3 central chip (target cluster excluded)
\tiny
\begin{tabular}{cccccccccccccc}

\hline
\hline
\multicolumn{14}{c}{\bf \small 3C295 Field}\\
\hline
\multicolumn{1}{c}{row} &
\multicolumn{1}{c}{ra$^a$ (2000)} &
\multicolumn{1}{c}{dec$^a$ (2000)} &
\multicolumn{1}{c}{Net Counts} &
\multicolumn{1}{c}{Bgd Counts} &
\multicolumn{1}{c}{Flux} &
\multicolumn{1}{c}{$\theta_{\rm off}$} &
\multicolumn{1}{c}{$B$} &
\multicolumn{1}{c}{$\alpha_{\rm ox}^b$} &
\multicolumn{1}{c}{z} &
\multicolumn{1}{c}{$L_{\rm X}^c$} &
\multicolumn{1}{c}{$M_{B}$} &
\multicolumn{1}{c}{class} &
\multicolumn{1}{c}{Ref.$^d$} \\
\multicolumn{1}{c}{} &
\multicolumn{1}{c}{} &
\multicolumn{1}{c}{} &
\multicolumn{1}{c}{} &
\multicolumn{1}{c}{} &
\multicolumn{1}{c}{10$^{-15}$cgs} &
\multicolumn{1}{c}{($^\prime$)} &
\multicolumn{1}{c}{} &
\multicolumn{1}{c}{} &
\multicolumn{1}{c}{} &
\multicolumn{1}{c}{10$^{42}$erg/s} &
\multicolumn{1}{c}{} &
\multicolumn{1}{c}{} &
\multicolumn{1}{c}{} \\
\hline
\hline
\multicolumn{14}{c}{\bf \small (0.5-2 keV)}\\
 1 &  14:11:23.4 & +52:13:31.4 & 360.33 $\pm$ 19.04 & 3.50 $\pm$ 0.02 & 55.85 $\pm$ 2.95 & 1.5 & 19.8 & 1.34&  0.47 &  58.6 & -22.6 & Sey1 & 4\\ 
 2 &  14:11:04.1 & +52:17:56.0 & 121.50 $\pm$ 11.59 & 2.56 $\pm$ 0.02 & 18.83 $\pm$ 1.80 & 6.3 & 18.6 & 1.73& & - & -23.6 & &\\ 
 3 &  14:11:19.5 & +52:14:00.0 & 112.82 $\pm$ 10.85 & 4.50 $\pm$ 0.02 & 17.49 $\pm$ 1.68 & 1.9 & 22.0 & 1.20  & 1.29 & 172.0 & -22.8 & QSO & 5 \\ 
 4 &  14:11:24.8 & +52:19:23.9 & 82.36 $\pm$ 9.65 & 2.11 $\pm$ 0.02 & 12.77 $\pm$ 1.50 & 7.4 & & & & - &  & &\\ 
 5 &  14:11:27.4 & +52:11:31.2 & 51.48 $\pm$ 7.38 & 3.25 $\pm$ 0.02 & 7.98 $\pm$ 1.14 & 1.4 & 20.1 & 1.62 & 0.59 & 13.9 & -22.8 & Sbc & 6\\ 
 6 &  14:10:59.5 & +52:11:52.8 & 51.15 $\pm$ 7.31 & 1.06 $\pm$ 0.01 & 7.93 $\pm$ 1.13 & 3.1 & & & & - &  & &\\ 
 7 &  14:11:30.8 & +52:14:23.4 & 35.40 $\pm$ 6.12 & 1.31 $\pm$ 0.01 & 5.49 $\pm$ 0.95 & 2.9 & 13.7& 2.68 & & - & -28.5 & star? &\\ 
 8 &  14:11:13.7 & +52:13:40.4 & 33.12 $\pm$ 5.92 & 1.92 $\pm$ 0.02 & 5.13 $\pm$ 0.92 & 1.8 & 21.2 & 1.53 & 2.08 & 147.0 & -24.8 & QSO & 4 \\ 
 9 &  14:11:09.0 & +52:16:45.1 & 30.34 $\pm$ 5.79 & 1.28 $\pm$ 0.01 & 4.70 $\pm$ 0.90 & 4.9 & & & & - & &  &\\ 
10 &  14:10:57.4 & +52:11:30.0 & 24.73 $\pm$ 5.16 & 1.01 $\pm$ 0.01 & 3.83 $\pm$ 0.80 & 3.5 & & & & - &  & &\\ 
11 &  14:11:14.5 & +52:16:10.9 & 22.44 $\pm$ 5.00 & 1.48 $\pm$ 0.01 & 3.48 $\pm$ 0.78 & 4.2 & & & & - &  & &\\ 
12 &  14:11:20.7 & +52:14:11.5 & 22.06 $\pm$ 4.93 & 2.08 $\pm$ 0.02 & 3.42 $\pm$ 0.76 & 2.1 & & & & - &  & &\\ 
13 &  14:11:29.2 & +52:13:32.5 & 20.11 $\pm$ 4.72 & 2.02 $\pm$ 0.02 & 3.12 $\pm$ 0.73 & 2.1 & 21.5 & 1.56 & 0.66 & 7.0 & -21.7 & Scd & 6 \\ 
14 &  14:11:26.3 & +52:18:47.8 & 16.33 $\pm$ 4.36 & 1.01 $\pm$ 0.01 & 2.53 $\pm$ 0.68 & 6.8 & & & & - &  & &\\ 
15 &  14:11:19.8 & +52:15:46.3 & 12.28 $\pm$ 3.73 & 1.13 $\pm$ 0.01 & 1.90 $\pm$ 0.58 & 3.7 & & & & - &  &  &\\ 
16 &  14:11:33.1 & +52:11:02.6 & 10.93 $\pm$ 3.51 & 1.20 $\pm$ 0.01 & 1.69 $\pm$ 0.54 & 2.4 & & & & - &  & &\\ 
17 &  14:11:27.0 & +52:14:14.0 & 8.24 $\pm$ 3.20 & 1.87 $\pm$ 0.01 & 1.28 $\pm$ 0.50 & 2.4 & & & & - &  & &\\ 
\multicolumn{14}{c}{\bf \small (2-10 keV)}\\
 1 &   14:11:23.4 & +52:13:31.2 & 80.04 $\pm$ 9.14 & 3.69 $\pm$ 0.02 & 91.2 $\pm$ 10.4  & & & & & 98.5 & & &\\ 
 2 &  14:11:04.1 & +52:17:55.8 & 45.95 $\pm$ 7.87 & 10.49 $\pm$ 0.04 & 52.4 $\pm$ 9  & & & & & - &  & &\\ 
 5 &  14:11:27.4 & +52:11:31.3 & 30.31 $\pm$ 5.80 & 3.53 $\pm$ 0.02 & 34.6 $\pm$ 6.6  & & & & & 60.6 & &  &\\ 
 3 &  14:11:19.4 & +52:14:00.4 & 29.40 $\pm$ 5.93 & 5.69 $\pm$ 0.03 & 33.5 $\pm$ 6.8  & & & & & 329.1 & &  &\\ 
\hline
\hline
\end{tabular}
\end{center}
\small
$^a$ Position uncertainties are estimated to be within $\sim$2''

$^b$ $\alpha_{\rm ox}=-log(S_{2 \rm keV}/S_{2500\AA})/2.605$, as defined in Stocke et al., 
1991

$^c$ z=0.5 and z=0.46 are assumed for RXJ0030 and 3C295 respectively, unless the redshift is 
available from the literature.

$^d$ References: (1) Ciliegi et al. 1995, (2) Boyle et al. 1995, (3) Brandt et al. 2000, 
(4) Dressler \& Gunn 1992, (5) Hewitt \& Burbidge 1989, (6) Thimm et al. 1994.
\end{table}

\begin{table}[htb]
\begin{center}
\small
{\bf Table A - 3:} list of point sources detected between 0.5-2 keV and 2-10 keV \\
in the anti-Leonid field - whole FOV (4 chips) \\
\tiny
\begin{tabular}{ccccccc}
\hline
\hline
\multicolumn{7}{c}{\bf \small anti-Leonid (4 chips)}\\
\hline
\multicolumn{1}{c}{row} &
\multicolumn{1}{c}{ra$^a$ (2000)} &
\multicolumn{1}{c}{dec$^a$ (2000)} &
\multicolumn{1}{c}{Net Counts} &
\multicolumn{1}{c}{Bgd Counts} &
\multicolumn{1}{c}{Flux} &
\multicolumn{1}{c}{$\theta_{\rm off}$} \\
\multicolumn{1}{c}{} &
\multicolumn{1}{c}{} &
\multicolumn{1}{c}{} &
\multicolumn{1}{c}{} &
\multicolumn{1}{c}{} &
\multicolumn{1}{c}{10$^{-15}$cgs} &
\multicolumn{1}{c}{($^\prime$)} \\
\hline
\hline
\multicolumn{7}{c}{\bf \small (0.5-2 keV)}\\
 1 &  22:13:52.4 & -22:15:56.2 & 333.20 $\pm$ 19.86 & 5.38 $\pm$ 0.03 & 82.97 $\pm$ 4.95 & 11.0 \\ 
 2 &  22:13:23.2 & -22:07:24.5 & 257.09 $\pm$ 16.37 & 1.20 $\pm$ 0.01 & 64.02 $\pm$ 4.08 & 6.2 \\ 
 3 &  22:12:57.5 & -22:21:35.5 & 213.74 $\pm$ 16.10 & 4.92 $\pm$ 0.02 & 53.22 $\pm$ 4.01 & 4.8 \\ 
 4 &  22:13:13.1 & -22:04:25.5 & 109.38 $\pm$ 10.90 & 1.18 $\pm$ 0.01 & 27.23 $\pm$ 2.71 & 4.5 \\ 
 5 &  22:12:49.1 & -22:11:32.2 & 74.91 $\pm$ 9.01 & 0.93 $\pm$ 0.01 & 18.65 $\pm$ 2.24 & 4.1  \\ 
 6 &  22:12:55.8 & -22:10:04.6 & 72.03 $\pm$ 8.69 & 0.67 $\pm$ 0.01 & 17.94 $\pm$ 2.16 & 5.5  \\ 
 7 &  22:12:40.1 & -22:07:48.1 & 51.28 $\pm$ 7.61 & 0.83 $\pm$ 0.01 & 12.77 $\pm$ 1.90 & 3.8 \\ 
 8 &  22:13:26.2 & -22:05:50.5 & 49.22 $\pm$ 7.39 & 1.87 $\pm$ 0.01 & 12.26 $\pm$ 1.84 & 2.3 \\ 
 9 &  22:12:25.1 & -22:07:28.5 & 43.80 $\pm$ 7.50 & 3.62 $\pm$ 0.02 & 10.91 $\pm$ 1.87 & 2.4 \\ 
 10 &  22:13:28.7 & -22:11:51.6 & 34.99 $\pm$ 6.08 & 0.43 $\pm$ 0.01 & 8.71 $\pm$ 1.51 & 1.4 \\ 
 11 &  22:13:19.5 & -22:08:36.0 & 31.99 $\pm$ 5.72 & 0.37 $\pm$ 0.01 & 7.97 $\pm$ 1.42 & 5.9 \\ 
 12 &  22:13:33.1 & -22:10:03.8 & 27.81 $\pm$ 5.48 & 0.64 $\pm$ 0.01 & 6.92 $\pm$ 1.36 & 1.4 \\ 
 13 &  22:12:43.4 & -22:18:38.1 & 24.44 $\pm$ 5.45 & 0.95 $\pm$ 0.01 & 6.09 $\pm$ 1.36 & 3.1 \\ 
 14 &  22:12:51.5 & -22:13:48.7 & 24.07 $\pm$ 5.16 & 0.65 $\pm$ 0.01 & 5.99 $\pm$ 1.29 & 5.3 \\ 
 15 &  22:12:54.2 & -22:19:05.3 & 23.68 $\pm$ 5.49 & 2.30 $\pm$ 0.02 & 5.90 $\pm$ 1.37 & 4.6 \\ 
 16 &  22:13:25.9 & -22:16:44.6 & 22.24 $\pm$ 5.03 & 0.73 $\pm$ 0.01 & 5.54 $\pm$ 1.25 & 3.2 \\ 
 17 &  22:13:37.9 & -22:08:29.3 & 22.14 $\pm$ 4.95 & 0.49 $\pm$ 0.01 & 5.51 $\pm$ 1.23 & 4.8 \\ 
 18 &  22:12:58.1 & -22:13:59.9 & 21.72 $\pm$ 4.85 & 0.49 $\pm$ 0.01 & 5.41 $\pm$ 1.21 & 3.1 \\ 
 19 &  22:12:47.7 & -22:17:49.8 & 19.10 $\pm$ 4.96 & 2.23 $\pm$ 0.02 & 4.76 $\pm$ 1.24 & 3.3 \\ 
 20 &  22:13:27.6 & -22:14:24.6 & 16.62 $\pm$ 4.26 & 0.42 $\pm$ 0.01 & 4.14 $\pm$ 1.06 & 5.1 \\ 
 21 &  22:13:41.0 & -22:15:10.1 & 16.06 $\pm$ 4.32 & 0.61 $\pm$ 0.01 & 4.00 $\pm$ 1.08 & 1.2 \\ 
 22 &  22:13:04.5 & -22:14:36.6 & 15.44 $\pm$ 4.08 & 0.35 $\pm$ 0.01 & 3.84 $\pm$ 1.02 & 3.9 \\ 
 23 &  22:13:00.5 & -22:08:32.9 & 14.00 $\pm$ 3.85 & 0.39 $\pm$ 0.01 & 3.49 $\pm$ 0.96 & 4.8 \\ 
 24 &  22:13:30.2 & -22:03:20.3 & 13.76 $\pm$ 4.01 & 0.64 $\pm$ 0.01 & 3.43 $\pm$ 1.00 & 6.1 \\ 
 25 &  22:13:12.6 & -22:08:52.5 & 13.62 $\pm$ 3.72 & 0.32 $\pm$ 0.01 & 3.39 $\pm$ 0.93 & 5.6 \\ 
 26 &  22:13:29.0 & -22:07:55.5 & 13.30 $\pm$ 3.78 & 0.33 $\pm$ 0.01 & 3.31 $\pm$ 0.94 & 8.0 \\ 
 27 &  22:12:54.5 & -22:08:01.4 & 13.24 $\pm$ 3.80 & 0.44 $\pm$ 0.01 & 3.30 $\pm$ 0.95 & 8.6 \\ 
 28 &  22:13:03.7 & -22:08:04.2 & 12.89 $\pm$ 3.69 & 0.40 $\pm$ 0.01 & 3.21 $\pm$ 0.92 & 8.0 \\ 
 29 &  22:12:57.5 & -22:19:15.9 & 12.55 $\pm$ 4.12 & 2.00 $\pm$ 0.02 & 3.13 $\pm$ 1.02 & 8.3 \\ 
 30 &  22:13:20.6 & -22:11:31.0 & 11.83 $\pm$ 3.50 & 0.28 $\pm$ 0.01 & 2.95 $\pm$ 0.87 & 8.2 \\ 
 31 &  22:13:07.9 & -22:15:25.9 & 10.30 $\pm$ 3.36 & 0.32 $\pm$ 0.01 & 2.56 $\pm$ 0.84 & 5.5 \\ 
 32 &  22:12:38.1 & -22:06:34.9 & 9.72 $\pm$ 3.56 & 1.37 $\pm$ 0.01 & 2.42 $\pm$ 0.89 & 7.3 \\ 
 33 &  22:13:08.1 & -22:07:22.9 & 8.89 $\pm$ 3.07 & 0.33 $\pm$ 0.01 & 2.21 $\pm$ 0.77 & 7.2 \\ 
 34 &  22:13:06.5 & -22:07:26.5 & 7.83 $\pm$ 2.90 & 0.37 $\pm$ 0.01 & 1.95 $\pm$ 0.72 & 9.9 \\ 
 35 &  22:13:49.1 & -22:11:12.2 & 7.57 $\pm$ 2.97 & 0.25 $\pm$ 0.01 & 1.88 $\pm$ 0.74 & 9.6 \\ 
 36 &  22:13:45.6 & -22:11:57.0 & 7.35 $\pm$ 2.94 & 0.38 $\pm$ 0.01 & 1.83 $\pm$ 0.73 & 1.1 \\ 
 37 &  22:13:20.3 & -22:03:24.7 & 7.22 $\pm$ 2.90 & 0.42 $\pm$ 0.01 & 1.80 $\pm$ 0.72 & 1.1 \\ 
 38 &  22:13:06.9 & -22:05:56.7 & 6.96 $\pm$ 2.78 & 0.37 $\pm$ 0.01 & 1.73 $\pm$ 0.69 & 3.3\\ 
 39 &  22:13:20.8 & -22:07:40.8 & 6.79 $\pm$ 2.72 & 0.39 $\pm$ 0.01 & 1.69 $\pm$ 0.68 & 9.8 \\ 
 40 &  22:13:18.4 & -22:10:20.5 & 6.75 $\pm$ 2.63 & 0.21 $\pm$ 0.00 & 1.68 $\pm$ 0.66 & 8.9 \\ 
\multicolumn{7}{c}{\bf \small (2-10 keV)}\\
  2 &  22:13:23.2 & -22:07:24.3 & 58.91 $\pm$ 7.97 & 2.22 $\pm$ 0.02 & 7.19 $\pm$ 0.97 & \\ 
  3 &  22:12:57.4 & -22:21:34.7 & 58.48 $\pm$ 9.44 & 14.46 $\pm$ 0.04 & 7.13 $\pm$ 1.15 & \\ 
  4 &  22:13:13.0 & -22:04:25.4 & 43.04 $\pm$ 7.04 & 2.93 $\pm$ 0.02 & 5.25 $\pm$ 0.86 & \\ 
 40 &  22:13:18.4 & -22:10:20.6 & 36.33 $\pm$ 6.13 & 1.46 $\pm$ 0.01 & 4.43 $\pm$ 0.75 & \\ 
  5 &  22:12:49.1 & -22:11:32.8 & 30.31 $\pm$ 5.95 & 2.65 $\pm$ 0.02 & 3.70 $\pm$ 0.73 & \\ 
  12 & 22:13:33.1 & -22:10:03.1 & 15.04 $\pm$ 4.35 & 2.75 $\pm$ 0.02 & 1.83 $\pm$ 0.53 & \\ 
 23 & 22:12:54.6 & -22:08:00.9  & 14.51 $\pm$ 4.08 & 1.23 $\pm$ 0.01 & 1.77 $\pm$ 0.50 & \\ 
  6 & 22:12:57.4 & -22:21:34.7 & 12.70 $\pm$ 3.89 & 1.78 $\pm$ 0.01 & 1.55 $\pm$ 0.47 & \\ 
\hline
\hline
\end{tabular}
\end{center}
\end{table}

\begin{table}[htb]
\begin{center}
\small
{\bf Table A - 4:} list of point sources detected between 0.5-2 keV and 2-10 keV \\
in the 3C273 field -- ACIS-S3 central chip\\
\tiny
\begin{tabular}{ccccccc}
\hline
\hline
\multicolumn{7}{c}{\bf \small 3C273 Field (S3)}\\
\hline
\multicolumn{1}{c}{row} &
\multicolumn{1}{c}{ra$^a$ (2000)} &
\multicolumn{1}{c}{dec$^a$ (2000)} &
\multicolumn{1}{c}{Net Counts} &
\multicolumn{1}{c}{Bgd Counts} &
\multicolumn{1}{c}{Flux} &
\multicolumn{1}{c}{$\theta_{\rm off}$} \\
\multicolumn{1}{c}{} &
\multicolumn{1}{c}{} &
\multicolumn{1}{c}{} &
\multicolumn{1}{c}{} &
\multicolumn{1}{c}{} &
\multicolumn{1}{c}{10$^{-15}$cgs} &
\multicolumn{1}{c}{($^\prime$)} \\
\hline
\hline
\multicolumn{7}{c}{\bf \small (0.5-2 keV)}\\
1$^b$ & 12:29:07.0 & +02:03:07.9 & 1095 $\pm$ 109 & 402 $\pm$ 0.22 & 1336 $\pm$ 12.9 & 1.1 \\ 
2 & 12:28:59.5 & +02:10:50.9 & 119.71 $\pm$ 11.84 & 7.64 $\pm$ 0.03 & 14.60 $\pm$ 1.44 & 7.1\\ 
3 & 12:29:15.4 & +02:05:29.5 & 99.75 $\pm$ 10.35 & 4.12 $\pm$ 0.02 & 12.17 $\pm$ 1.26 & 3.4\\ 
4 & 12:29:07.2 & +02:04:00.8 & 31.52 $\pm$ 6.32 & 9.05 $\pm$ 0.03 & 3.85 $\pm$ 0.77 & 0.9\\ 
5 & 12:29:08.5 & +02:05:53.9 & 30.10 $\pm$ 5.72 & 2.25 $\pm$ 0.02 & 3.67 $\pm$ 0.70 & 2.4\\ 
6 & 12:29:14.5 & +02:01:21.2 & 18.06 $\pm$ 4.75 & 3.59 $\pm$ 0.02 & 2.20 $\pm$ 0.58 & 3.8\\ 
7 & 12:28:59.3 & +02:05:28.3 & 17.57 $\pm$ 4.60 & 3.49 $\pm$ 0.02 & 2.14 $\pm$ 0.56 & 1.9\\ 
8 & 12:29:02.2 & +02:05:24.9 & 16.53 $\pm$ 4.46 & 3.42 $\pm$ 0.02 & 2.02 $\pm$ 0.54 & 1.6\\ 
9 & 12:28:52.2 & +02:05:13.2 & 16.49 $\pm$ 4.47 & 2.92 $\pm$ 0.02 & 2.01 $\pm$ 0.55 & 3.1\\ 
10 & 12:29:11.1 & +02:05:31.0 & 15.29 $\pm$ 4.30 & 2.92 $\pm$ 0.02 & 1.87 $\pm$ 0.52 & 2.5\\ 
11 & 12:28:52.2& +02:05:13.2 & 14.11 $\pm$ 4.24 & 2.70 $\pm$ 0.02 & 1.72 $\pm$ 0.52 & 5.2\\ 
12 & 12:29:17.4& +02:08:31.6 & 11.81 $\pm$ 4.15 & 3.99 $\pm$ 0.02 & 1.44 $\pm$ 0.51 & 5.9\\ 
13 & 12:29:18.0& +02:06:10.8 & 9.93 $\pm$ 3.62 & 2.51 $\pm$ 0.02 & 1.21 $\pm$ 0.44 & 4.3\\ 
\multicolumn{7}{c}{\bf \small (2-10 keV)}\\
1$^b$ & 12:29:07.5 & +02:03:08.1 & 4299 $\pm$ 68.1 & 251 $\pm$ 0.17 & 4084 $\pm$ 63.4 & \\
 4 & 12:29:07.3& +02:04:01.7 & 60.18 $\pm$ 8.88 & 20.01 $\pm$ 0.05 & 57.17 $\pm$ 8.44 & \\ 
 3 & 12:29:15.4& +02:05:30.0 & 35.04 $\pm$ 6.35 & 3.98 $\pm$ 0.02 & 33.28 $\pm$ 6.03 & \\ 
 - & 12:29:11.7& +02:03:13.3 & 26.98 $\pm$ 6.30 & 12.24 $\pm$ 0.04 & 25.63 $\pm$ 5.98 & \\ 
 2 & 12:28:58.6& +02:10:53.1 & 10.55 $\pm$ 3.95 & 3.37 $\pm$ 0.02 & 10.02 $\pm$ 3.75 & \\ 
\hline
\end{tabular}
\end{center}
\small
\hspace{4cm}
$^a$ Position uncertainties are estimated to be within $\sim$2''

$^b$ Source (3C273) affected by pile-up, so values should be regarded as only indicative. 
\end{table}

\end{document}